\documentclass[twocolumn]{aastex63}
\usepackage{multirow}
\usepackage{color}
\usepackage{lineno}

\usepackage{graphicx}
\usepackage{float}
\usepackage{amsmath,amssymb}
\usepackage{xcolor}
\usepackage{natbib}
\usepackage[hang,flushmargin]{footmisc}

\usepackage{hyperref}
\usepackage{enumitem}
\setlist{parsep=0pt,listparindent=\parindent}

\newcommand{\ntotalsne}{916}
\newcommand{\ndiscreports}{778}

\newcommand{\percentdisc}{8.3}
\newcommand{\nspecclass}{138}

\usepackage{pifont}

\newcommand{\Cambridge}{Institute of Astronomy and Kavli Institute for Cosmology, Madingley Road, Cambridge, CB3 0HA, UK}
\newcommand{\JHU}{Department of Physics and Astronomy, The Johns Hopkins University, Baltimore, MD 21218.}
\newcommand{\STScI}{Space Telescope Science Institute, Baltimore, MD 21218.}

\newcommand{\IfA}{Institute for Astronomy, University of Hawaii, 2680 Woodlawn Drive, Honolulu, HI 96822, USA}

\newcommand{\UCSC}{Department of Astronomy and Astrophysics, University of California, Santa Cruz, CA 95064, USA}
\newcommand{\QUB}{Astrophysics Research Centre, School of Mathematics and Physics, Queen's University Belfast, Belfast BT7 1NN, UK}

\newcommand{\Einstein}{NASA Einstein Fellow}
\newcommand{\Hubble}{Hubble Fellow}

\newcommand{\Northwestern}{Center for Interdisciplinary Exploration and Research in Astrophysics (CIERA) and Department of Physics and Astronomy, Northwestern University, Evanston, IL 60208, USA}
\newcommand{\DARK}{DARK, Niels Bohr Institute, University of Copenhagen, Jagtvej 128, 2200 Copenhagen, Denmark}
\newcommand{\Illinois}{Department of Astronomy, University of Illinois at Urbana-Champaign, 1002 W. Green St., IL 61801, USA}
\newcommand{\NCSA}{Center for Astrophysical Surveys, National Center for Supercomputing Applications, Urbana, IL, 61801, USA}
\newcommand{\Toronto}{David A. Dunlap Department of Astronomy and Astrophysics, University of Toronto, 50 St. George Street, Toronto, Ontario, M5S 3H4 Canada}
\newcommand{\WSU}{Department of Physics \& Astronomy, Washington State University, Pullman, Washington 99164, USA}

\newcommand{\NCUG}{Graduate Institute of Astronomy, National Central University, 300 Zhongda Road, Zhongli, Taoyuan 32001, Taiwan}
\newcommand{\ESO}{European Southern Observatory, Alonso de C{\'o}rdova 3107, Vitacura, Santiago, Chile}
\newcommand{\Melbourne}{School of Physics, The University of Melbourne, VIC 3010, Australia}
\newcommand{\astrothreed}{ARC Centre of Excellence for All Sky Astrophysics in 3 Dimensions (ASTRO 3D)}
\newcommand{\Southhampton}{Department of Physics and Astronomy, University of Southampton, Highfield, Southampton SO17 1BJ, UK}
\newcommand{\HKU}{Department of Physics, The University of Hong Kong, Pokfulam Road, Hong Kong, China}
\newcommand{\Milan}{Dipartimento di Fisica, Universit\`a  degli Studi di Milano, via Celoria 16, I-20133 Milano, Italy}
\newcommand{\Sternberg}{Sternberg Astronomical Institute, Lomonosov Moscow State University 13 Universitetsky pr., Moscow 119234, Russia}
\newcommand{\Carnegie}{Observatories of the Carnegie Institute for Science, 813 Santa Barbara St., Pasadena, CA 91101, USA}

\newcommand{\RateUnit}{{\rm yr}^{-1}{\rm Mpc}^{-3}}

\begin{document}

\title{The Young Supernova Experiment: Survey Goals, Overview, and Operations}

\suppressAffiliations

\author[0000-0002-6230-0151]{D.~O.~Jones}
\altaffiliation{\Einstein}
\affiliation{\UCSC}


\author[0000-0002-2445-5275]{R.~J.~Foley}
\affiliation{\UCSC}

\author[0000-0001-6022-0484]{G.~Narayan}
\affiliation{\Illinois}
\affiliation{\NCSA}

\author[0000-0002-4571-2306]{J.~Hjorth}
\affiliation{\DARK}

\author[0000-0003-1059-9603]{M.~E.~Huber}
\affiliation{\IfA}


\author[0000-0002-6298-1663]{P.~D.~Aleo}
\affiliation{\Illinois}
\affiliation{\NCSA}

\author[0000-0002-8297-2473]{K.~D.~Alexander}
\altaffiliation{\Einstein}
\affiliation{\Northwestern}

\author[0000-0002-4269-7999]{C.~R.~Angus}
\affiliation{\DARK}

\author[0000-0002-4449-9152]{K.~Auchettl}
\affiliation{\Melbourne}
\affiliation{\astrothreed}
\affiliation{\DARK}
\affiliation{\UCSC}

\author[0000-0003-4703-7276]{V.~F.~Baldassare}
\affiliation{\WSU}

\author[0000-0002-5369-6094]{S.~H.~Bruun}
\affiliation{\DARK}

\author[0000-0001-6965-7789]{K.~C.~Chambers}
\affiliation{\IfA}

\author[0000-0003-0038-5468]{D.~Chatterjee}
\affiliation{\Illinois}
\affiliation{\NCSA}

\author[0000-0001-5126-6237]{D.~L.~Coppejans}
\affiliation{\Northwestern}

\author[0000-0003-4263-2228]{D.~A.~Coulter}
\affiliation{\UCSC}

\author[0000-0003-4587-2366]{L.~DeMarchi}
\affiliation{\Northwestern}

\author[0000-0001-9494-179X]{G.~Dimitriadis}
\affiliation{\UCSC}

\author[0000-0001-7081-0082]{M.~R.~Drout}
\affiliation{\Toronto}

\author[0000-0003-2348-483X]{A.~Engel}
\affiliation{\Illinois}
\affiliation{\NCSA}

\author[0000-0002-4235-7337]{K.~D.~French}
\altaffiliation{\Hubble}
\affiliation{\Illinois}
\affiliation{\NCSA}
\affiliation{\Carnegie}

\author[0000-0003-4906-8447]{A.~Gagliano}
\affiliation{\Illinois}
\affiliation{\NCSA}

\author[0000-0002-8526-3963]{C.~Gall}
\affiliation{\DARK}

\author[0000-0002-9878-7889]{T.~Hung}
\affiliation{\UCSC}

\author[0000-0001-9695-8472]{L.~Izzo}
\affiliation{\DARK}

\author[0000-0002-3934-2644]{W.~V.~Jacobson-Gal\'an}
\affiliation{\Northwestern}

\author[0000-0002-5740-7747]{C.~D.~Kilpatrick}
\affiliation{\UCSC}

\author[0000-0003-0529-1161]{H.~Korhonen}
\affiliation{\ESO}
\affiliation{\DARK}

\author[0000-0003-4768-7586]{R.~Margutti}
\affiliation{\Northwestern}

\author[0000-0002-6248-398X]{S.~I.~Raimundo}
\affiliation{\Southhampton}
\affiliation{\DARK}

\author[0000-0003-2558-3102]{E.~Ramirez-Ruiz}
\affiliation{\UCSC}
\affiliation{\DARK}

\author[0000-0002-4410-5387]{A.~Rest}
\affiliation{\JHU}
\affiliation{\STScI}

\author[0000-0002-7559-315X]{C.~Rojas-Bravo}
\affiliation{\UCSC}

\author{M.~R.~Siebert}
\affiliation{\UCSC}

\author[0000-0002-8229-1731]{S.~J.~Smartt}
\affiliation{\QUB}

\author[0000-0001-9535-3199]{K.~W.~Smith}
\affiliation{\QUB}

\author[0000-0003-0794-5982]{G.~Terreran}
\affiliation{\Northwestern}

\author[0000-0001-5233-6989]{Q.~Wang}
\affiliation{\JHU}

\author[0000-0001-9666-3164]{R.~Wojtak}
\affiliation{\DARK}

\author[0000-0001-9775-0331]{A.~Agnello}
\affiliation{\DARK}

\author[0000-0002-4775-9685]{Z.~Ansari}
\affiliation{\DARK}

\author[0000-0001-5409-6480]{N.~Arendse}
\affiliation{\DARK}

\author{A.~Baldeschi}
\affiliation{\Northwestern}

\author[0000-0003-0526-2248]{P.~K.~Blanchard}
\affiliation{\Northwestern}

\author[0000-0001-6415-0903]{D.~Brethauer}
\affiliation{\Northwestern}

\author[0000-0002-7735-5796]{J.~S.~Bright}
\affiliation{\Northwestern}

\author{J.~S.~Brown}
\affiliation{\UCSC}

\author[0000-0001-5486-2747]{T.~J.~L.~de~Boer}
\affiliation{\IfA}

\author[0000-0002-3696-8035]{S.~A.~Dodd}
\affiliation{\UCSC}

\author[0000-0002-2833-2344]{J.~R.~Fairlamb}
\affiliation{\IfA}

\author[0000-0002-5926-7143]{C.~Grillo}
\affiliation{\Milan}
\affiliation{\DARK}

\author[0000-0003-2349-101X]{A.~Hajela}
\affiliation{\Northwestern}

\author[0000-0001-7666-1874]{C.~Hede}
\affiliation{\DARK}

\author[0000-0001-7364-4964]{A.~N.~Kolborg}
\affiliation{\DARK}

\author[0000-0001-8825-4790]{J.~A.~P.~Law-Smith}
\affiliation{\UCSC}

\author[0000-0002-7272-5129]{C.-C.~Lin}
\affiliation{\IfA}

\author[0000-0002-7965-2815]{E.~A.~Magnier}
\affiliation{\IfA}

\author[0000-0001-7179-7406]{K.~Malanchev}
\affiliation{\Illinois}
\affiliation{\Sternberg}

\author[0000-0002-4513-3849]{D.~Matthews}
\affiliation{\Northwestern}

\author[0000-0001-6350-8168]{B.~Mockler}
\affiliation{\UCSC}
\affiliation{\DARK}

\author[0000-0002-5788-9280]{D.~Muthukrishna}
\affiliation{\Cambridge}

\author[0000-0001-8415-6720]{Y.-C.~Pan}
\affiliation{\NCUG}

\author[0000-0003-0841-5182]{H.~Pfister}
\affiliation{\DARK}
\affiliation{\HKU}

\author[0000-0002-6029-7163]{D.~K.~Ramanah}
\affiliation{\DARK}

\author[0000-0002-3825-0553]{S.~Rest}
\affiliation{\JHU}

\author[0000-0002-9820-679X]{A.~Sarangi}
\affiliation{\DARK}

\author[0000-0003-1735-8263]{S.~L.~Schr\o der}
\affiliation{\DARK}

\author[0000-0001-8769-4591]{C.~Stauffer}
\affiliation{\Northwestern}

\author[0000-0002-3019-4577]{M.~C.~Stroh}
\affiliation{\Northwestern}

\author[0000-0002-5748-4558]{K.~L.~Taggart}
\affiliation{\UCSC}

\author[0000-0002-1481-4676]{S.~Tinyanont}
\affiliation{\UCSC}

\author[0000-0002-1341-0952]{R.~J.~Wainscoat}
\affiliation{\IfA}

\collaboration{1000}{(Young Supernova Experiment)}

\correspondingauthor{D.~O.~Jones}
\email{david.jones@ucsc.edu}
  
\begin{abstract}
Time domain science has undergone a revolution over the past decade, with tens of thousands of new supernovae (SNe) discovered each year. However, several observational domains, including SNe within days or hours of explosion and faint, red transients, are just beginning to be explored.  Here, we present the Young Supernova Experiment (YSE), a novel optical time-domain survey on the Pan-STARRS telescopes. Our survey is designed to obtain well-sampled $griz$ light curves for thousands of transient events up to $z \approx 0.2$. This large sample of transients with 4-band light curves will lay the foundation for the Vera C.\ Rubin Observatory and the {\it Nancy Grace Roman Space Telescope}, providing a critical training set in similar filters and a well-calibrated low-redshift anchor of cosmologically useful SNe~Ia to benefit dark energy science.  As the name suggests, YSE complements and extends other ongoing time-domain surveys by discovering fast-rising SNe within a few hours to days of explosion. YSE is the only current four-band time-domain survey and is able to discover transients as faint $\sim$21.5~mag in $gri$ and $\sim$20.5~mag in $z$, depths that allow us to probe the earliest epochs of stellar explosions. YSE is currently observing approximately 750 square degrees of sky every three days and we plan to increase the area to 1500 square degrees in the near future. When operating at full capacity, survey simulations show that YSE will find $\sim$5000 new SNe per year and at least two SNe within three days of explosion per month. To date, YSE has discovered or observed \percentdisc\% of the transient candidates reported to the International Astronomical Union in 2020. We present an overview of YSE, including science goals, survey characteristics and a summary of our transient discoveries to date.
\end{abstract}


\section{Introduction}
\label{sec:intro}

For thousands of years, astrophysical transients were discovered only by chance.  It was \citet{Zwicky38} who began the first systematic search for extragalactic astrophysical transients, which evolved into the Palomar Supernova Search, discovering over 100 supernovae (SNe) in the following few decades.  Systematic searches of the Southern sky started in the 1980s \citep{Maza80}, essentially doubling capabilities.  Combining charge-coupled devices (CCDs) with robotic telescopes to automatically search for SNe was the next major innovation \citep{Kare81}, although successful searches started about a decade after initial tries \citep{Perlmutter89, Filippenko92, Richmond93}.

The discovery of SN~1987A \citep{Kunkel87} caused a resurgence of interest in SN science and at a similar time, advances in calibrating SNe~Ia, building upon work of \citet{Kowal68}, \citet{Rust74}, and \citet{Pskovskii77}, indicated that SNe~Ia would be precise distance indicators capable of measuring the expansion history of the Universe \citep{Phillips93}.  The innovative Calan/Tololo survey \citep{Hamuy93}, which used photometric plates for discovery and CCDs for follow-up observations, was key to a significant increase in SN discovery.  Quickly, pencil-beam surveys designed to discover high-redshift SNe began providing the majority of discoveries \citep{Norgaard-Nielsen89, Perlmutter97, Schmidt98}.  Rolling searches, where the search epochs and follow-up observations are combined, became an efficient method for observing many transients \citep{Barris04}, and we still use this strategy today.

In the late 1990s, the Lick Observatory Supernova Search \citep[LOSS;][]{Filippenko05} and extremely sophisticated amateur astronomers\footnote{\url{http://www.rochesterastronomy.org/snimages/lindex.html}} \citep[e.g.,][]{Evans94} continued to increase the discovery rate of nearby SNe.  Additional scientific discoveries such as the detection of SN progenitor stars in pre-explosion images \citep[e.g.,][]{Woosley87, Aldering94}, the connection between SNe and long-duration gamma-ray bursts \citep{Galama98, Hjorth03, Stanek03}, and the discovery of the accelerating expansion of the Universe \citep{Riess98, Perlmutter99} made transient discovery even more valuable.

In the last two decades, the discovery rate of astrophysical transients has been increasing at an exponential rate\footnote{SN discovery statistics can be found at \url{https://wis-tns.weizmann.ac.il/stats-maps}, \url{http://www.rochesterastronomy.org/sn2020/snstats.html} and from the Open Supernova Catalog \citep{Guillochon17}.}.  This is mainly the result of systematic searches for high-redshift SNe (Supernova Cosmology Project, \citealt{Perlmutter97}; High-Z Supernova Search, \citealt{Schmidt98}; Deep Lens Survey, \citealt{Becker04}; Supernova Legacy Survey, \citealt{Astier06, Guy10}; ESSENCE, \citealt{Miknaitis07, Narayan16}; SDSS-II, \citealt{Frieman08, Kessler09, Sako18}, Dark Energy Survey, \citealt{Bernstein12, DES16, Abbott19, Brout19}; Pan-STARRS Medium-Deep Survey, \citealt{Rest14, Jones18, Villar20}; {\it HST} surveys CANDELS, CLASH, and the Frontier Fields, \citealt{Graur_2014, Rodney14, Kelly15}, Subaru Hyper Suprime-Cam Transient Survey, \citealt{Tanaka16}) and low-redshift SNe (SNFactory, \citealt{Aldering02}; Texas Supernova Search, \citealt{Quimby06}; SkyMapper, \citealt{Keller07, Scalzo17}; Catalina Real-Time Transient Survey, \citealt{Drake09}; (i)PTF, \citealt{Law09}; CHASE, \citealt{Pignata09}; MASTER, \citealt{Lipunov10}; ATLAS, \citealt{Tonry11, Tonry18}; La Silla QUEST, \citealt{Baltay13}; ASAS-SN, \citealt{Shappee14}; PSST, \citealt{Huber15}; DLT40, \citealt{Valenti17}; ZTF, \citealt{Bellm19}).  Recent low-redshift all-sky surveys such as ATLAS, ASAS-SN and PSST in particular were critical to increasing the transient discovery rate in the last few years, with ZTF now augmenting the rate even further.  This recent wealth of time domain data has led to significant advancement in our understanding of stellar evolution, SN explosion mechanisms, black holes, the chemical enrichment of galaxies and the fundamental physics of our universe \citep{Campana06, Lorimer07, Smith07, Smartt09, Nomoto13, Riess16, Abbott17, Scolnic18}.  In the coming decade, the Vera C.\ Rubin Observatory will further increase the rate of transient discoveries by {\it an order of magnitude}, potentially finding $10^{5}$ new transients per year \citep{LSST09}.

A number of smaller-scale time domain surveys are complementing these efforts with multi-wavelength and fast-cadence transient searches.  Surveys including the Vista Infrared Extragalactic Legacy Survey \citep{Honig16} and the GALEX time domain survey \citep{Gezari13} undertook transient searches in the near-infrared and ultraviolet, respectively.  Current fast-cadence searches, including the ZTF one-day survey, DLT40 \citep{Valenti17}, the Evryscope \citep{Law15}, the Korea Microlensing Telescope Network \citep{Kim16}, Kepler (K2; \citealp{Howell14}) and TESS \citep{Fausnaugh19} have pioneered new techniques to understand transient and variable phenomena on short timescales.  However, a number of key questions remain unanswered by current surveys due to their limited wavelength coverage, area, cadence, depth, or photometric calibration.

Here, we describe the Young Supernova Experiment (YSE), a three-year survey for transients that focuses on discovering statistical samples of young transients with additional emphasis on discovering rare and red transients, measuring cosmological parameters, and preparing for the Rubin Observatory.  YSE began on November 24th, 2019, and is currently using 7\% of the observing time on Pan-STARRS1 to survey 750~deg$^2$ of sky with a 3-day cadence to a depth of $gri \approx 21.5$~mag and $z \approx 20.5$~mag; we plan to double the survey area to $\sim$1500~deg$^2$ in the near future.  Our survey strategy emphasizes increased coverage in $iz$ and improved depth to distinguish our transient discovery demographics from other ongoing surveys. It will also leverage the excellent photometric calibration of PS1 for SN~Ia cosmology.  When possible, we attempt to interleave our observations with those of ZTF for an alternating one- or two-day combined cadence, further improving our ability to identify young transients.  

Below, in Section \ref{sec:science}, we discuss key open questions and challenges in transient astrophysics that can be addressed through a new wide-angle time domain survey such as YSE $-$ in particular, understanding SN progenitors through observations of young supernovae, building a census of faint, fast, and red transients, measuring cosmological parameters, understanding black hole variability and tidal disruption events (TDEs), and preparing for Rubin Observatory science.  
We will use these goals to motivate the YSE survey strategy in Section~\ref{sec:strategy}.  YSE vetting and follow-up procedures are described in Section~\ref{sec:broker} and an overview of the YSE survey status and discoveries to date are described in Section~\ref{sec:yields}.

\section{YSE Science Drivers: Open Questions and Challenges in Transient Astrophysics}
\label{sec:science}

\subsection{Young Supernovae}

In the hours after a stellar explosion, a SN is still roughly the size of its progenitor star.  The shock breakout, which produces a large X-ray and UV photon flux, will ``flash ionize'' the immediate circumstellar medium \citep[CSM; e.g.,][]{Gal-Yam14}.  During this time, the SN shock will also interact with nearby CSM and any potential companion star, ablating material from its surface.  As the ejecta expands, it will smooth out the initial inhomogeneities caused by asymmetric explosions and the initial conditions of the progenitor star.

Flash ionization provides a unique window to examine the CSM before the SN ejecta sweeps it up.  If the ionizing spectrum is known and the CSM is optically thin, one can convert measured fluxes from lines of different atomic species into densities and abundances at different radii.  Combined with multi-wavelength observations \citep[e.g.,][]{Margutti11fe, Chomiuk11fe}, one can create a holistic picture of the progenitor star's circumstellar environment \citep{Jacobson-Galan20}.

Any radioactive elements in the outermost layers of the SN ejecta will also result in additional flux before energy produced in the deeper layers of the ejecta has time to diffuse out.  As a result, this can produce ``excess'' flux relative to the later rising light curve \citep{Piro13}.   Excess flux has been seen for several SNe~Ia \citep{Marion16, Hosseinzadeh17, Dimitriadis19, Shappee19, Miller20} and other peculiar thermonuclear transients \citep{Cao15, Jiang17}, but either because of a limited number of data points or limited color information, the interpretation is often unclear \citep[e.g.,][]{Dimitriadis19b, Tucker19}.  Earlier detections, higher cadence, additional color information, and early spectroscopy could break such degeneracies.

Luminous mass-loss episodes {\it prior} to the SN explosion may also be detectable given a survey with sufficient depth and cadence (\citealp{Ofek14}; for a review, see \citealp{2014ARA&A..52..487S}).  A particularly spectacular example was a luminous outburst two years before the explosion of SN~Ibn~2006jc \citep{Foley07, Pastorello07}. Slightly more common are outbursts likely associated with luminous blue variables (LBVs) before a terminal explosion, such as in SN~2009ip \citep{Smith10:09ip, Foley11b, 2013ApJ...767....1P,2014ApJ...780...21M} and 2015bh \citep{Elias-Rosa16,Thone17}.  Early bumps in the light curves of superluminous SNe (SLSNe) have also been found \citep[e.g.,][]{Leloudas12,Nicholl15,Smith16,Nicholl17}, and may be due to outbursts prior to the explosion or interaction between the SN and shells of CSM.  A survey with deep time-domain imaging would provide enormous legacy value for analysis of future transients --- perhaps discovered even decades later --- by enabling deep searches for such outbursts.

\subsection{Rare, Faint, Fast, and Red Transients}

There is considerable uncertainty in our understanding of the nature of rare classes and subclasses of transients, including peculiar thermonuclear SNe (e.g., SN Iax; SN~Ia-CSM; \citealp{Hamuy03}; \citealt{Foley13:iax}; Ca-rich SNe; \citealt{Perets10}; SN~2000cx-like; \citealp{Li01}; SN~2002es-like; \citealt{Ganeshalingam10}; SN~2006bt-like; \citealt{Foley10}; various He-shell explosions \citealt{De19, Jacobson-Galan20b, Miller20, Siebert20}), superluminous supernovae \citep[e.g.,][]{Gal-Yam12}, tidal disruption events \citep{Gezari13}, low-luminosity transients \citep[e.g.,][]{Foley09, Valenti09, Rodney18, Shubham20}, fast-evolving blue optical transients (FBOTs; \citealp{Drout14}), LBV outbursts \citep[e.g.,][]{VanDyk00, Kilpatrick18}, and kilonovae \citep{Coulter17, Drout17, Tanvir17}.

Transients from many of these classes are rarely discovered because they are intrinsically low luminosity, fast evolving, or red \citep{Siebert17}.  
Traditional surveys such as LOSS \citep{Leaman11} were designed to detect common SNe and so had cadences of 5--15~days with limiting magnitudes set to detect SNe~II in targeted galaxies.  These surveys were usually performed with blue-sensitive filters.  Such a survey would therefore miss many transients of these uncommon classes even if the intrinsic volumetric rate was high. Surveys such as Gattini-IR \citep{De2020}, with $J$-band observations every two days and a median limiting mag $J = 15.7$, are just now beginning to explore a redder discovery space.
Transients such as intermediate luminosity red transients (ILRTs; \citealp{Bond09}) and luminous red novae (LRN; \citealp[e.g.,][]{Mould90,Mason10, Nicholls13}), associated with stellar outbursts and  mergers, are some of the faint transient classes that may have been frequently missed by previous surveys.

Additionally, these small field of view (FoV) surveys targeted luminous galaxies to increase their overall discovery yield and rejected potential transients at the centers of galaxies to avoid subtraction false positives and active galactic nucleus (AGN) activity.  However some rare classes occur preferentially in low-luminosity galaxies \citep{Lunnan15} and nuclear regions \citep{Bloom11}, and previous surveys would miss these transients even if they were luminous, blue, and long-lived.

A new, untargeted, red-sensitive survey would not only increase rare transient discoveries from individual epochs but would have the potential to increase discoveries of faint but long-duration transients by stacking multiple search epochs obtained over a period of several nights.   
This would have a significant impact, for example, on the discovery of gravitationally lensed SNe (glSNe) and superluminous SNe (SLSNe), where the SNe are luminous and long-lived but their volumetric rates are low.  Both classes are predominantly found at high redshift and extend the cosmological eras that can be probed spectroscopically.  Red bands are advantageous for glSN discoveries in particular \citep{Wojtak2019} and systematic detections of glSNe could pave the way for constraining the Hubble constant (H$_0$) from gravitational time delay measurements \citep{Refsdal1964}.  For SLSNe \citep{Knop99,Smith10,Quimby11}, many questions remain about their progenitors, explosion mechanisms, luminosity sources, intrinsic rates, interaction with circumstellar material, and how feedback from these sources affects their host galaxies \citep{Gal-Yam19}.  Surveys capable of discovering SLSNe before peak with high-cadence multi-band light curves will enable rapid identification, providing sources for multi-wavelength follow-up and a large statistical sample.

\subsection{Cosmology}

Perhaps counter-intuitively, it is the nearest SN~Ia samples that are responsible for most of the systematic uncertainty on measurements of the dark energy equation-of-state parameter, $w$ \citep{Brout19b, Jones19}.  Most existing low-$z$ SN\,Ia observations were compiled on more than 13 different, though partially correlated, photometric systems at a time when cosmological analyses were not yet limited by mmag-level uncertainties.  It is critical to replace these legacy data with large, well-calibrated samples of low-$z$ SNe\,Ia for next-generation cosmological analyses. Large statistical samples of low-$z$ SNe are also ideal for measuring the growth of structure from their peculiar velocities to test general relativity \citep{Howlett17, Huterer2017, Boruah20, Kim19} and reducing the statistical uncertainty of H$_0$ \citep{Riess16}.

The 2020s will have two groundbreaking facilities for measuring cosmological parameters with SNe~Ia: the Rubin Observatory and the {\it Roman Space Telescope}.  However, neither will provide an optimal low-$z$ sample for cosmological parameter measurements.  {\it Roman} will find few low-$z$ SNe \citep{Hounsell18}, while the nominal-cadence strategy of Rubin will create large ($\sim$20-day) single-filter gaps in their low-$z$ light curves that would make SN\,Ia standardization less accurate.  Therefore, cosmological measurements with either observatory will likely rely on external low-$z$ data sets.  Large, high-cadence, unbiased low-$z$ samples could also be used to refine and extend new models for SN\,Ia standardization \citep[e.g.,][]{Guy07,Burns11,Saunders18,Leget20,Mandel20}, to better constrain relationships between SN\,Ia distance measurements and host galaxy properties \citep{Rigault13,Jones18b,Roman18,Rigault18}, to improve our understanding of common sources of CC\,SN contamination in SN\,Ia cosmology measurements \citep{Jones18,Popovic20,Vincenzi20}, and to increase the number of independent photometric systems on which SNe are observed in order to reduce photometric calibration uncertainties --- typically the dominant systematic uncertainty in dark energy measurements.

\subsection{Black Hole Variability and Tidal Disruption Events}

Varying accretion rates in active galactic nuclei (AGN) and QSOs are reflected in their optical variability.  By discovering and monitoring a large number of quasars, it is possible to revisit tests of the damped-random walk model of quasar variability \citep{Kelly09,MacLeod10} and study correlations between quasar variability and physical properties \citep[e.g.,][]{Sanchez-Saez18, Kimura20}. Optical variability is also a useful tool for identifying the relatively elusive AGN in low-mass galaxies \citep[e.g.,][]{Butler11, Baldassare18, Sanchez-Saez19, Baldassare20}.

A small fraction of these AGN and QSOs will evolve dramatically over a period of a few years \citep{MacLeod16, Ruan16}, changing their brightness, hardness of their spectrum, ionization state, and presence of broad spectral lines \citep{LaMassa15, Runnoe16}.  Prompt spectroscopic observations immediately after photometric changes are detected may then reveal the underlying physics of these ``changing-look quasars'' \citep[e.g.,][]{Macleod19}.  By discovering changing-look AGN in the early stages of their transition, one can monitor the change in black hole accretion physics and the possible build up of an accretion disc \citep{Gezari17}.
 
A survey that can reliably find transients in the cores of galaxies and identify them early also allows TDE discoveries.  Detailed multi-wavelength follow-up observations of TDEs can trace their full evolution, including the formation of accretion disks \citep{Hung20}, the formation of jets \citep{Zauderer11}, and the properties of the black holes themselves \citep[e.g.,][]{Auchettl17, Hung17, Holoien19, Mockler19, Alexander20, vanVelzen20}. Studying the diversity of TDEs constrains the black hole mass function and extends it to lower masses than is possible with AGN observations alone \citep{MacLeod12, Kochanek2016, French20}.
Observations of AGN or TDEs from {\it wandering} black holes could constrain their dynamics, a key phase before the formation of black hole binaries, and track the ways in which SMBHs settle to the centers of their host galaxies \citep{2019MNRAS.482.2913B, 2020ApJ...888...36R}.

\subsection{Preparation for the Rubin Observatory Era}

One of the key transient science challenges of the forthcoming Rubin Observatory era is how to identify unusual or scientifically valuable transients from samples of hundreds of thousands.  This could include classifications within the first days or hours of discovery, allowing subsequent spectroscopic or photometric follow-up observations, or full light-curve classifications, necessary for a census of SN rates and luminosity functions as well as for cosmological parameter measurements with SNe~Ia.  Recent efforts to classify a diverse sample of transients have been successful \citep[e.g.,][]{Boone19, Muthukrishna19, Villar19, Moller20} including those that exclusively use contextual host galaxy information \citep{Foley13, Baldeschi20, Gagliano20}, but the sensitivity of these classifiers to biases and non-representative training samples is often unclear.  Training data sets built from existing transient surveys will typically have significantly different photometric filters or cadences than Rubin.  Multi-color light curves for thousands of transients within a magnitude-limited but otherwise unbiased discovery space would serve as a vital test bed for training transient brokers and designing efficient follow-up strategies.  Light curves with $iz$ coverage in particular are missing from most current time-domain surveys.
SNe from the Pan-STARRS medium deep survey will assist with these goals \citep{Hosseinzadeh20,Villar20}, but even that sample lacks examples of several rare classes and rarely has detections within a few days of explosion.

Building a survey that coordinates observations between multiple telescopes would also be useful logistical and conceptual preparation for the Rubin Observatory era, which will pre-announce its pointing plan.  This will allow supplemental observations from other telescopes that improve the effective cadence of Rubin.

\subsection{Magnitude- and Volume-limited Census of SNe}

In spite of thousands of new SN discoveries per year, the rates and luminosity functions of many CC~SN classes --- especially fainter and redder classes --- are highly uncertain \citep{Li11,Perley20}. This propagates to uncertainty in photometric classification of SNe \citep{Jones17}, stellar evolution, the physics of SN explosions, and the chemical enrichment of the Universe \citep{Maoz17}.  The most often used local rate measurements were measured from a galaxy-targeted survey \citep{Li11}, and therefore the rates and luminosity functions are heavily biased by the correlations between SNe and their galaxy environments \citep{Smith07, Quimby11, Sanders12, Taggart19}.  Blue-sensitive surveys will also not discover highly reddened transients, resulting in a systematic uncertainty for their rate.  Since some SNe, particularly those with short-lived progenitor stars, will preferentially occur in dusty environments, the level of bias will be different for each class.

A number of teams, including ASAS-SN \citep{Holoien19} and the ZTF Bright Transient Survey \citep{Fremling20}, are carrying out large surveys to spectroscopically classify all transients within a given magnitude for an improved census of transients that does not rely on galaxy associations, redshift catalogs, and assumed distances.  However, a survey with greater sensitivity at redder wavelengths would increase the sample of rare, red SNe that can be discovered and an extended volume-limited survey, similar to, e.g., \citet{De2020b} from ZTF, 
would ensure that low-luminosity transients are well represented in the measured rates and luminosity functions.  Combining magnitude- and volume-limited surveys leverages the statistics of luminous events in the magnitude-limited sample while preserving the low-luminosity transients in the volume-limited sample.  A combined sample would constrain the stellar initial mass function and the progenitor systems of transient classes and subclasses \citep[][]{Strolger15}.

\begin{figure*}
    \centering
    \includegraphics[width=7in]{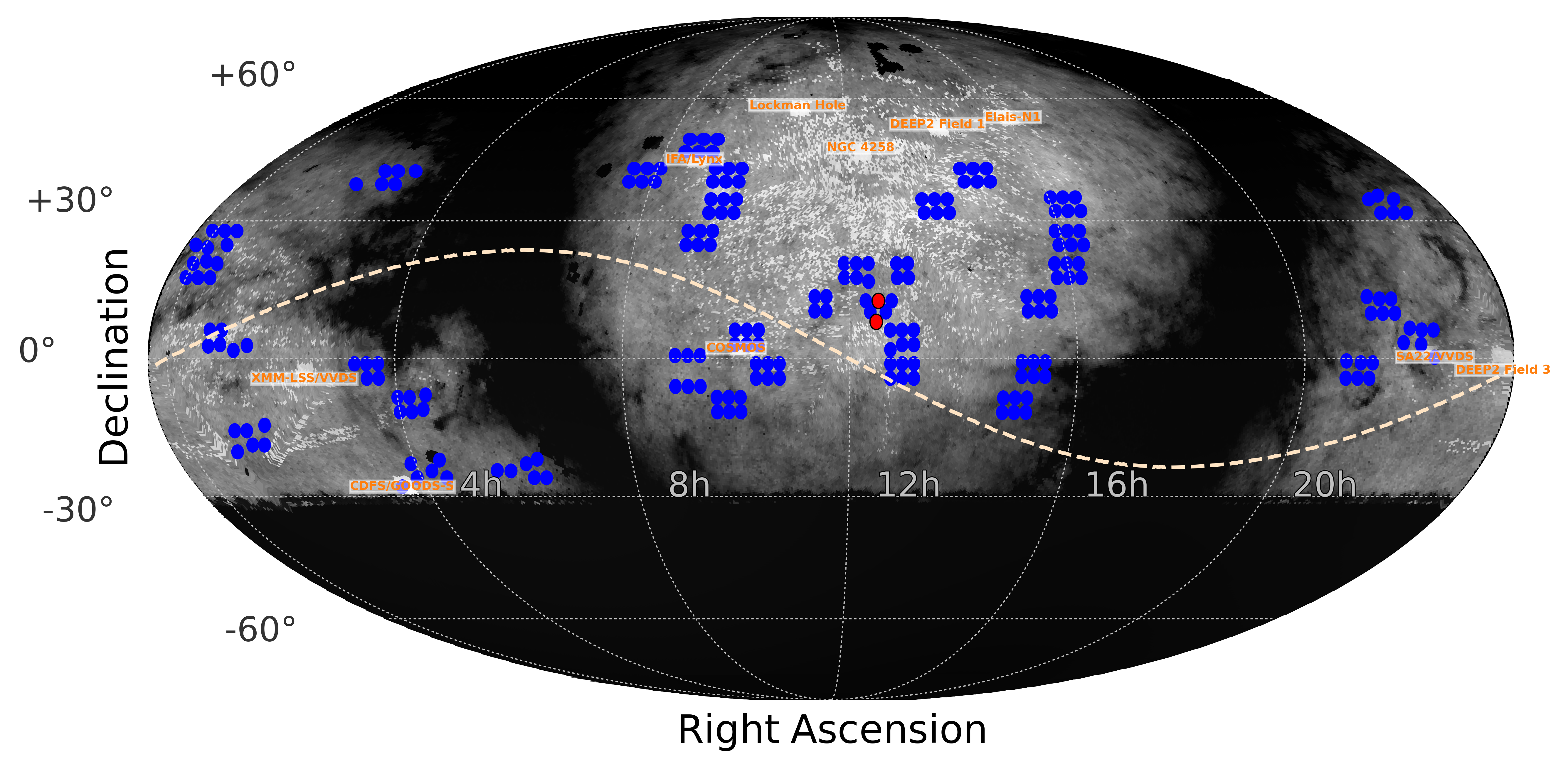}
    \caption{YSE fields chosen prior to October 1$^{\rm st}$, 2020 (blue).  The red fields correspond to the two daily Virgo pointings. Lighter (white) background regions correspond to a higher value for the YSE field selection metric (Section~\ref{sec:fieldsel}).  In addition to this metric, we typically require fields to be at least 20 degrees from the ecliptic plane (dashed line).  The labeled high-metric white squares are at the locations of the Pan-STARRS medium deep fields.}
    \label{fig:initial_fields}
\end{figure*}

\section{The Young Supernova Experiment: Survey Overview and Strategy}
\label{sec:strategy}

\begin{table}
\caption{\fontsize{9}{11}\selectfont YSE Survey Characteristics}
  \centering
\begin{tabular}{lr}
  \hline \hline\\[-1.5ex]

Area$^{\mathrm{a}}$&1512 deg$^2$\\
Cadence&3 days\\
Exposure Time&27s\\
Filter Sequence (dark)&$gr$, $gi$\\
Filter Sequence (bright)&$ri$, $rz$\\
Med.\ $gri$ Depth (dark)$^{\mathrm{b}}$&$21.52, 21.65, 21.37$~mag\\
Med.\ $riz$ Depth (bright)$^{\mathrm{b}}$&$20.87, 20.93, 20.50$~mag\\
Med.\ FWHM&1.3\arcsec\\
Med.\ cadence$^{\mathrm{c}}$&3.9 days\\
Pixel scale&0.25\arcsec pix$^{-1}$\\
Total area&$\sim$7000~deg$^2$\\
$griz$ stack depths$^{\mathrm{d}}$&$23.6,23.7,23.6,22.8$~mag\\

\hline\\[-1.5ex]
\multicolumn{2}{l}{
\begin{minipage}{7.5cm}
$^{\mathrm{a}}$ Due to detector masking, YSE will survey approx.\ 1200~deg$^2$ per epoch, with a variable position angle allowing coverage of the full 1500~deg$^2$ over a given observing season.  Current YSE observations use PS1 only and cover $\sim$750~deg$^2$, but we will soon double the area by commencing PS2 observations.

$^{\mathrm{b}}$ 5$\sigma$ depths are computed by injecting artificial sources in the nightly YSE images and testing how many are recovered.

$^{\mathrm{c}}$ Between March 1$^{\rm st}$ and October 1$^{\rm st}$, 2020 to exclude large gaps due to telescope downtime in the first months of the survey.  Light curve gaps due to telescope position angle changes for some transients are not included.

$^{\mathrm{d}}$ Estimated five-$\sigma$ depths for one-year YSE stacks calculated using the depth computation from \citet{Chambers16} (see Sections \ref{sec:depth} and \ref{sec:science_impact}) and assuming 30\% of epochs are lost due to weather and 24\% are lost due to detector masking.  Over a subset of the YSE area, $\sim 2000$~deg$^2$, we will reach depths of $griz \approx 24.1,24.3,24.2,23.5$~mag by combining three years of YSE data.  See Table~\ref{table:stacks} for details.

\end{minipage}}
\end{tabular}
\label{table:yse_overview}

\end{table}

\begin{table*}
\caption{\fontsize{9}{11}\selectfont Large-Area, Extragalactic Transient Surveys}
  \centering
\begin{tabular}{lrrrrrrrrrrr}
  \hline \hline\\[-1.5ex]
  Survey & Area$^\mathrm{a}$ & Cadence & Filters & Mag Lim.$^{\mathrm{b}}$ & \multicolumn{2}{c}{$M = -19$ Volume$^{\mathrm{c}}$} & \multicolumn{2}{c}{$M = -16 $ Volume$^{\mathrm{c}}$} & Pixel & Calib.\\
  &&&&& all&low-$A_V$ &  all &low-$A_V$ & Scale &\\
  & (deg$^{2}$)& (days) &&& \multicolumn{2}{c}{($10^{-3}$~Gpc$^{3}$)} & \multicolumn{2}{c}{($10^{-3}$~Gpc$^{3}$)} & $(\arcsec/\mathrm{pix}$) & (mmag)\\
\hline\\[-1.5ex]
ATLAS&24500&2&$co^{\mathrm{d}}$&$c/o \approx 19.7$&296&230&5.97&4.63&1.86&5\\
\\
ASAS-SN$^{\mathrm{e}}$&20000&1&$Vg$&$V \approx 17^{\mathrm{f}}$&7&5&0.13&0.10&8&\nodata\\
\\
PSST$^{\rm g}$&14000&$\sim$0-15&$iw$&$iw \approx 21,22$&140&\nodata&2.29&\nodata&0.25&3\\
\\
ZTF\\
$-$ MSIP$^{\rm h}$&12975&3&$gr$&$gr = 21.1,20.9$&850&740&20.22&17.51&1&10\\ 
$-$ $i$-band&7900&4&$i$&$i = 20.2$&177&\nodata&3.75&\nodata&1&10\\
$-$ ZTF one-day&1725&1&$gr$&$gr = 21.1,20.9$&113&\nodata&2.69&\nodata&1&10\\
\\
\hline\\[-1.5ex]
YSE\\
$-$ Full Survey&1512&3&$griz$&$gr = 21.5,21.7$&161&161&4.07&4.07&0.25&3\\
$-$ $i$-band only&1512&6&$i$&$i = 21.4$&135&135&$3.35$&3.35&0.25&3\\
$-$ $z$-band only&1512&12&$z$&$z = 20.5^{\mathrm{i}}$&49&49&$1.07$&1.07&0.25&3\\
$-$ $+$ZTF MSIP&1512&1.5&$griz$&$gr_{\mathrm{ZTF}} = 21.1,20.9$&70&60&$1.59$&1.59&\nodata&5\\
\hline\\[-1.5ex]
\multicolumn{11}{l}{
  \begin{minipage}{18.5cm}
    $^{\mathrm{a}}$ Area per cadence cycle.

    $^{\mathrm{b}}$ Dark time, when available.

    $^{\mathrm{c}}$ Using the bluest available band, neglecting MW dust but restricting the volume calculation to approx.\ $E(B-V) < 0.2$ area in the low-$A_V$ column \citep{Schlafly11}.
    
    $^{\mathrm{d}}$ ``Cyan'' and ``Orange'' bands, combinations of $g+r$ and $r+i$, respectively.
    
    $^{\mathrm{e}}$ Or $\sim$30000~deg$^2$ every 2-3 days.
    
    $^{\mathrm{f}}$ Separate dark/bright time limits were not given by \citet{Holoien17},
    with the ASAS-SN website (\url{http://www.astronomy.ohio-state.edu/~assassin/index.shtml}) quoting limits down to 18th magnitude.
    
    $^{\mathrm{g}}$ The PSST cadence is irregular, with $\sim$1-4 return visits for a given field typically scheduled within 15 days after the initial observation.
    
    $^{\mathrm{h}}$ The ZTF Phase I survey; the recently begun ZTF Phase II MSIP survey is a two-day cadence all sky survey using 50\% of the total telescope time.
    
    $^{\mathrm{i}}$ Bright time limit as YSE does not observe in $z$ during dark time.

    {\bf Note.}  Selected extragalactic transient surveys sorted by area per cadence cycle.  Limiting magnitudes and volumes are computed for dark time observations. Approximate magnitude limits, survey area over the time period of one cadence cycle, and calibration are from \citet{Bellm19b,Bellm19,Masci19} for ZTF, from \citet{Schlafly12,Scolnic15} and this work for YSE, from M.\ Huber (private communication) for PSST, and from \citet{Tonry18} for ATLAS, and from \citet{Holoien17} for ASAS-SN.  Volume estimates assume flat $\Lambda$CDM with H$_0$ = 70~km~s$^{-1}$~Mpc$^{-1}$ and $\Omega_m = 0.3$. 
\end{minipage}}
\end{tabular}
\label{table:yse_comp}

\end{table*}

The science goals above motivate the YSE survey design, field selection and strategy: we wish to design a wide-angle, deep survey with multi-color light curves, including significant coverage at redder wavelengths, and with a survey strategy that facilitates obtaining a statistical sample of young SN detections.
Below, we give an overview of the Pan-STARRS telescopes and photometric system, discuss and motivate the YSE survey properties, and use survey simulations to estimate the SN yields from the chosen survey design.  YSE fields chosen to date are shown in Figure~\ref{fig:initial_fields} and basic survey characteristics are given in Table~\ref{table:yse_overview}.

\subsection{Overview of the Pan-STARRS Telescopes, Photometric System, and Data Processing Pipeline}

YSE observations use the two 1.8-meter Pan-STARRS telescopes (Pan-STARRS1 and Pan-STARRS2), each with 1.4 gigapixel cameras (GPC1 and GPC2; \citealt{Kaiser02}).  The Pan-STARRS telescopes have an approximately 7~deg$^2$ FoV, with that of Pan-STARRS2 (PS2) slightly larger due to additional CCD chips at the corners of the FoV and the GPC2 detector's reduced masking compared to GPC1.  Pan-STARRS1 (PS1) commenced formal survey operation in May of 2010 and has currently imaged over 3$\pi$ steradians of the sky \citep{Chambers16}.  PS2 recently finished commissioning.  There are sub-percent color transformations between the PS1 and PS2 photometric systems.

Pan-STARRS observations are taken through one of six broadband filters, $grizyw_{P1}$ (hereafter $grizyw$).  The filter transmissions and total system throughput for these filters have been measured via a calibrated photodiode and tunable laser \citep{Tonry12}.  Combined with the well-measured filter functions, the 3$\pi$~steradians of Pan-STARRS sky coverage allowed \citet{Schlafly12} to solve for the relative calibration of the PS1 photometric system to a precision of better than 5~mmag.  \citet{Scolnic15} then improved the absolute calibration of PS1 by comparing observations of secondary standard stars across multiple photometric systems.  The excellent photometric calibration of PS1 is critical for SN cosmology and helpful for many of the other YSE science goals discussed in Section \ref{sec:science}.

Pan-STARRS data are processed using the Image Processing Pipeline (IPP) at the University of Hawaii's Institute for Astronomy \citep{Magnier16}.  The IPP is used to download, process and archive all Pan-STARRS images, and includes a difference imaging pipeline to search for transients.  Once data are taken and reduced, and a template image is convolved and subtracted from the data, the Transient Science Server at Queens University Belfast (Section~\ref{sec:broker}; \citealp{Smith20}) uses a combination of massive catalog cross matching and a machine learning algorithm to discover transient phenomena in the survey images.  

The Pan-STARRS design and infrastructure has enabled a number of innovative transient science programs, notably the Medium Deep Survey (MDS), the Pan-STARRS survey for Transients (PSST; \citealp{Huber15}), and recently the Pan-STARRS Search for Kilonovae \citep{McBrien21}.  The Medium Deep Survey observed 70~deg$^2$ of sky at an average of six observations per ten days from 2010 to 2014, discovering over 5000 SNe.  Approximately 500 of these SNe were classified spectroscopically \citep{Rest14,Scolnic18} and spectroscopic host galaxy redshifts for over 3000 were obtained \citep{Jones17}.  Science highlights from these data include cosmological parameter measurements \citep{Rest14,Jones18}, tidal disruption event discoveries \citep{Gezari13, Chornock14}, fast-evolving luminous SNe \citep{Drout14}, SLSNe \citep{Chomiuk11,Lunnan13,McCrum14}, SNe\,Iax \citep{Narayan11} and transient classification tools for the Rubin Observatory \citep{Villar19}.

PSST commenced after the end of the MDS and used the data taken on PS1 through NASA's near earth object (NEO) observation program to discover new transients.  PSST combines the IPP difference imaging pipeline at the IfA in Hawaii and the Transient Science Server at Queen's University Belfast to search for transients.  The PSST team has discovered the majority of the $\sim$9000 publicly reported PS1 transients.  The Pan-STARRS Search for Kilonovae is using these same NEO observations to conduct a volume-limited search for intrinsically faint transients within $\sim$200~Mpc \citep{McBrien21}.  Though YSE will not carry out LIGO counterpart searches, Pan-STARRS is also a powerful facility for searching for gravitational wave counterparts, due to both the field of view and the existence of reference sky templates for immediate difference images above $\delta \approx -30$. Pan-STARRS has has been employed for this work during the LIGO Scientific Collaboration and Virgo Collaboration's observing runs O1--O3 \citep[e.g.,][]{Smartt16,Ackley20}.

\begin{figure}
    \centering
    \includegraphics[width=3.4in]{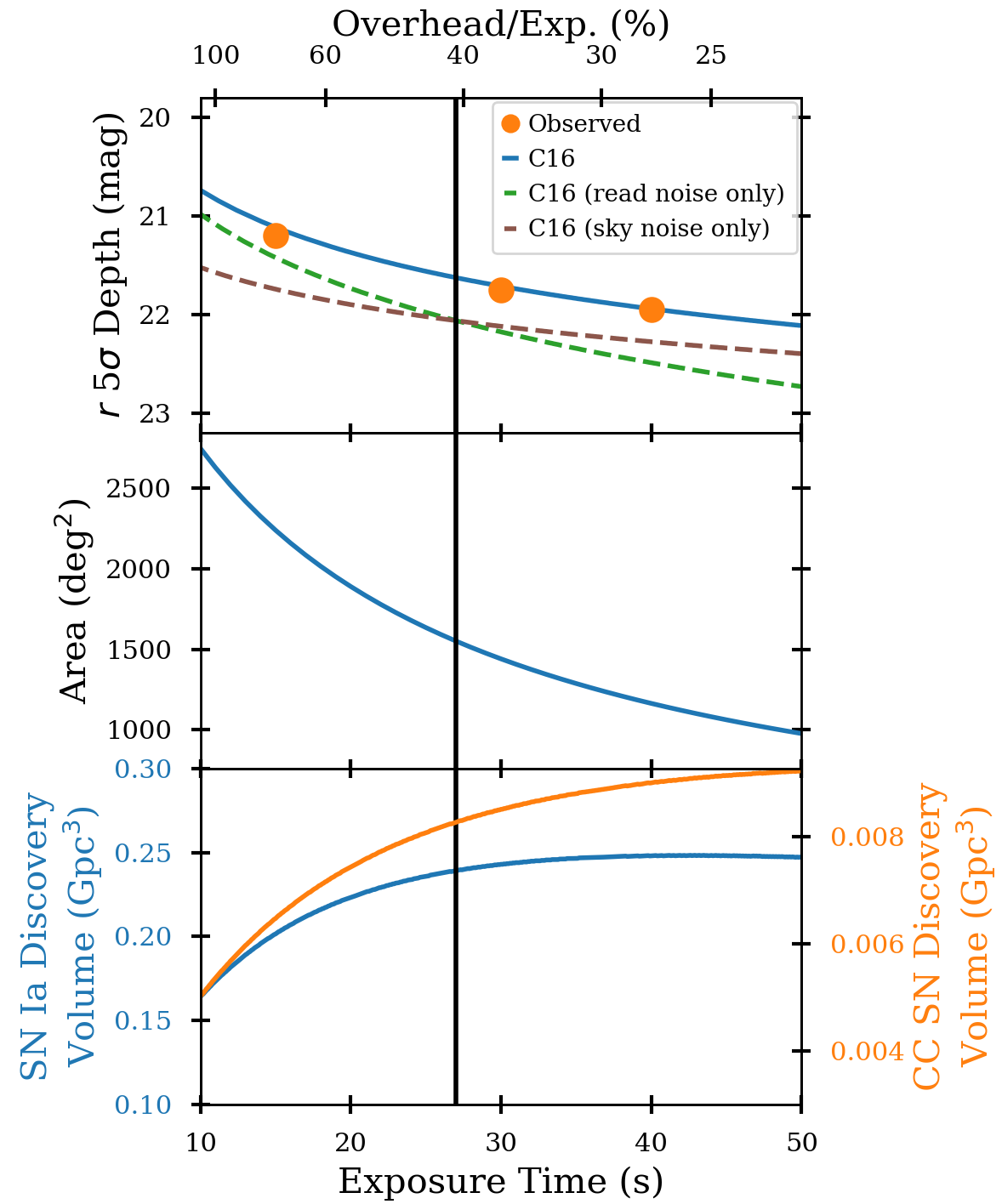}
    \caption{Top: Depth as a function of exposure time for PS1 with read noise (green), sky noise (brown), and total noise (blue).  Observed depths from the Foundation Supernova Survey at 15, 30, and 40~seconds are shown in orange with the overhead (11s) divided by the exposure time shown on the top axis.  Middle: YSE area as a function of exposure time for our fixed total observing time.  Bottom: YSE discovery volume for SNe~Ia assuming $M_{r} = -19.3$~mag (blue; left axis) and a nominal CC~SN with  $M_{r} = -16.5$~mag (orange; right axis) as a function of exposure time for our fixed total observing time.  The black vertical line indicates the chosen YSE exposure time (27~sec), which corresponds to the exposure time at which the sky noise begins to dominate in the $r$ band.}
    \label{fig:depth}
\end{figure}

\subsection{Depth, Cadence, Area, and Filters}
\label{sec:depth}

The Pan-STARRS telescopes allow for a large-area, high-cadence, well-calibrated survey, but place a number of constraints on the exposure time and area coverage of YSE.  Our team considered only exposure times greater than 15~sec; for practical purposes, this is the minimum allowed exposure time to limit systematic uncertainties in the photometry due to the GPC1 shutter.  However, the Pan-STARRS cameras also have a relatively high read noise of 8~e$^{-}$ \citep{Chambers16} and an overhead of approximately 11~sec per exposure, which makes it advantageous to increase the YSE exposure time beyond the 15-sec minimum.  

Figure~\ref{fig:depth} demonstrates the trade-off between area and volume for different PS1 exposure times.  The read noise and sky noise for PS1 during dark time are given by \citet{Chambers16}, and agree with measured depths from Foundation Supernova Survey data \citep{Foley18} and PSST.  We wish to survey a large volume while still covering a large enough area to discover many brighter, nearby events, and while surveying a significant fraction of the ZTF area\footnote{See also \citet{Bellm16} for quantitative discussion of the tradeoffs between survey volume, area, and cadence for maximizing detection rate of transients and evaluating discovery capabilities of transient surveys.}. Overlap with ZTF enables high-cadence light curves from combined YSE and ZTF data.

After evaluating a number of different strategies (Appendix \ref{sec:sim_appendix}), our team chose $griz$ exposure times of 27~sec.  This choice of exposure time satisfies a number of useful criteria.  First, YSE observations are $\sim$0.4--0.8~mag deeper than ZTF, enabling day-before observations --- both detections and non-detections --- that put useful limits on the ages of newly discovered SNe.  Second, while YSE data are read-noise dominated in $g$ observations taken during dark time, this exposure time ensures that the rest of the data will be background-dominated.  Finally, a 27-sec exposure time nearly maximizes the possible YSE survey volume (Figure~\ref{fig:depth}), but still covers a large area such that more than 100 SNe/year will have $r < 18.5$~mag, making them easy to follow spectroscopically using 2- and 3-meter-class telescopes from the ground.

We considered several different filter sequences for YSE, a number of which are discussed in Appendix~\ref{sec:sim_appendix}.  We decided on a strategy that emphasizes the $iz$ filters while also having at least one $g$ or $r$ band observation per epoch, as $gr$ observations allow us to measure rise times for young SNe from the combination of YSE and ZTF data without any assumptions on the color of the event.  YSE observes in only two filters per night to maximize the survey area while still preserving color information for transient observations.  During dark time, we alternate observations of $gr$ and $gi$, and in bright time, due to the $\sim$1~mag lower $g$-band depth, we alternate $ri$ and $rz$ observations.  During brief periods of grey time (moon illumination between 33\% and 66\%) we alternate $gi$ and $gz$.  

Finally, YSE observes each field with a three-day cadence to increase synergy with the ZTF observing strategy and to allow well-sampled light curves for relatively fast-evolving transients.

In Table~\ref{table:yse_comp} we compare the YSE survey characteristics to those of other ongoing time-domain surveys.  Each of these surveys has unique advantages, with ASAS-SN discovering nearby SNe with extremely fast cadence, ATLAS having a slightly slower cadence compared to ASAS-SN but increased depth and excellent calibration, and ZTF covering less area than ASAS-SN/ATLAS but having greater depth and both high-cadence and $i$-band sub-surveys.  In terms of area observed at a given time, YSE will observe significantly less area than ASAS-SN, ATLAS, and ZTF.  However, YSE's depth is 0.4-0.8~mag deeper than any other survey and therefore YSE will cover more volume than ASAS-SN, 50-70\% as much volume as ATLAS, depending on the transient luminosity, and $\sim$20\% as much volume as ZTF with the $i$-band data in particular probing a unique volume equal to 76-90\% of the ZTF $i$-band volume.  In terms of the photometric calibration in particular, the excellent PS1 calibration outperforms other systems (and PS2 calibration will be tied to PS1 for comparable accuracy), which is particularly important for SN\,Ia cosmology.  The over-sampled PS1 point spread function (PSF) may allow for improved image subtraction, which benefits black hole variability and TDE studies by facilitating discoveries near the centers of galaxies.  The combined ZTF/YSE cadence is particularly advantageous for fast-cadence studies, helping with discoveries of young SNe, and the $z$ band coverage is unique to YSE and will aide in the discovery of very red transients.

\begin{figure*}
    \centering
    \includegraphics[width=7in]{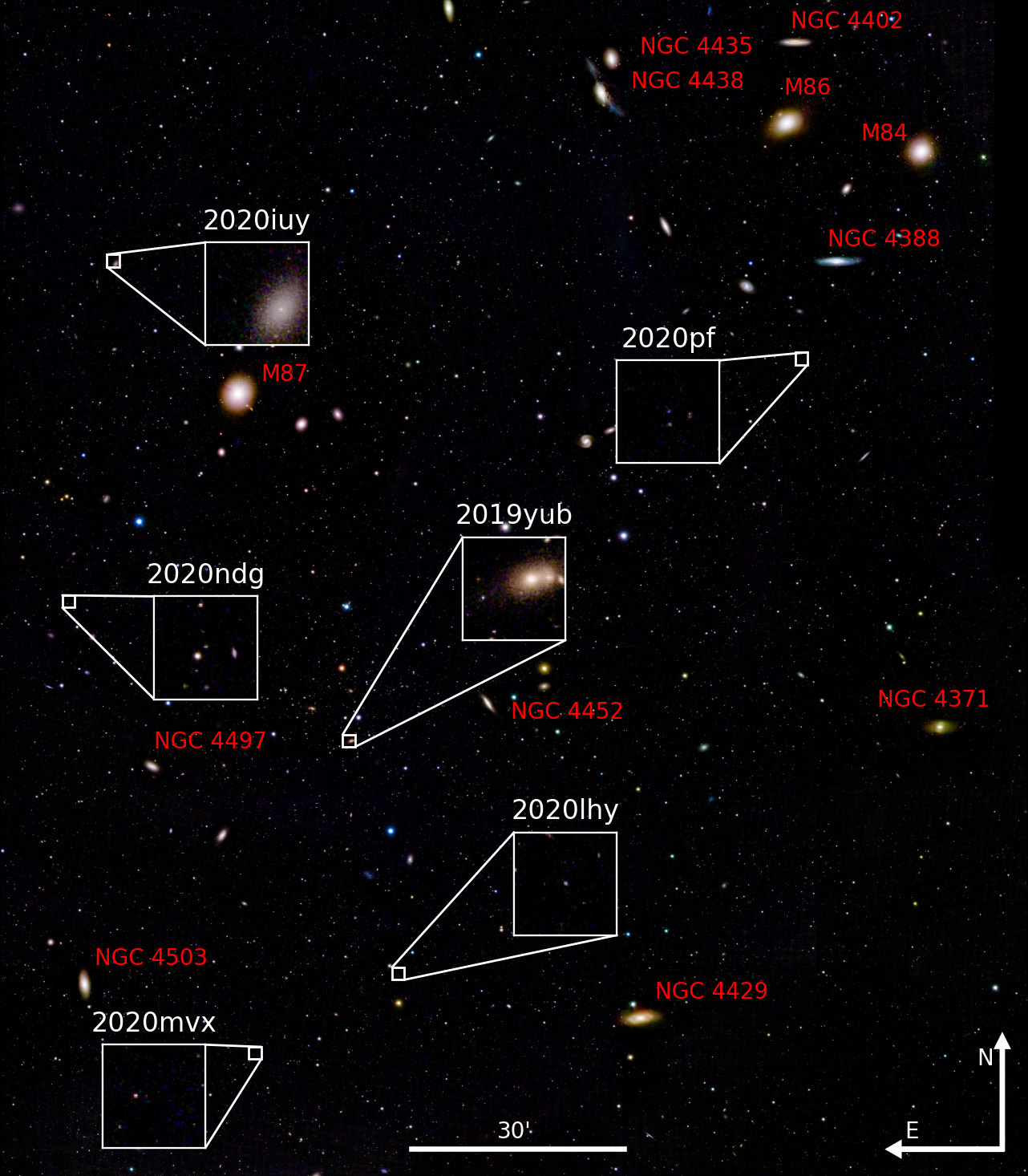}
    \caption{Stacked $2 \deg \times 2.4 \deg$ image corresponding to part of a YSE daily Virgo cluster field.  A subset of luminous Virgo cluster galaxies in the field are labeled in red.  White boxes show an enlarged view of the host galaxies of transients discovered or observed by YSE in this field, including likely stellar outburst AT~2020iuy at $M_i \approx -12$~mag.}
    \label{fig:virgo}
\end{figure*}

\subsection{Field Selection}
\label{sec:fieldsel}

We select YSE survey fields based on the following criteria:

\begin{itemize}
    \item To increase transient discovery volume, we choose fields with high Galactic latitude and low Milky Way extinction.
    \item We prioritize fields with a large amount of available archival data, primarily redshifts and deep multi-wavelength imaging.
    \item We prioritize equatorial fields, as they can be observed with follow-up facilities from both hemispheres. However, in most cases we require fields to be at least 20 degrees from the ecliptic plane to minimize observing gaps caused by the Pan-STARRS moon avoidance angle of 30 degrees.
    \item We avoid fields with declinations less than $-30$ degrees as YSE requires deep template images from previous Pan-STARRS observations, which are scarce at large negative declinations.
    \item We choose fields that, given their history of observations, ZTF will typically observe on a three-day cadence or greater and in two filters.
    \item Our observations are interwoven with Pan-STARRS near earth object observations, requiring us to space our fields in right ascension so that they can be observed throughout the night.
    \item When possible, we place pointings in minimum schedulable blocks of six pointings each, as Pan-STARRS slews of more than 15 degrees require approximately one minute of overhead to refocus the telescope.
    \item We prioritize new fields with rising or scientifically interesting transients that have been discovered by other time-domain surveys and for which YSE data would be beneficial.  We will attempt to follow such transients using as many fields as possible in the first year of the survey and using 50\% of fields in subsequent years.  The remaining 50\% will be kept the same across multiple years for long-term monitoring.
    \item We prioritize fields that have a larger number of nearby galaxies within 150~Mpc, with galaxies at $<$10~Mpc given the highest weight.  This will result in a slight bias to the nearby rates in exchange for an increase in SN discoveries but we expect to be able to correct for this bias in future analyses given the large fraction of galaxies that are untargeted.
\end{itemize}

\noindent We use a semi-arbitrary field selection metric to take these priorities into account, which is discussed in Appendix~\ref{sec:fields_appendix}.  We consider the total amount of {\it HST} exposure time when selecting fields, which is important for identifying or constraining progenitor systems in pre-explosion imaging \citep[e.g.,][]{Li11:11fe, McCully14, Foley15:14dt, Kilpatrick18:17ein, Kilpatrick18:17eaw}, as well as the number of active SNe --- particularly SNe before maximum light --- in a given field, but do not formally include these quantities in the metric.  The initial set of YSE pointings is shown in Figure~\ref{fig:initial_fields}.

\subsubsection{Daily Survey of Virgo}

When the Virgo Cluster is observable at ${\rm airmass} < 1.5$, YSE will also dedicate two Pan-STARRS pointings to a daily survey of Virgo.  Our team may choose to undertake similar mini-surveys of other nearby clusters, such as Coma, in the future.  The Virgo region of the sky produces an extraordinary number of SNe; in the last 15 years, 30 SNe were spectroscopically classified within the radii of the two YSE Virgo pointings our team has adopted\footnote{Using data collected by the Open Supernova Catalog; \citealp{Guillochon17}.} (not accounting for detector masking).  Most recently, this includes three SNe~Ia (SNe 2018bgb, 2018axs, and 2017eea) and one SN~IIn (SN2017jfs).

YSE will be able to detect pre-explosion outbursts for Virgo transients to an approximate absolute magnitude of $-10$.  We have chosen our pointings to maximize the archival {\it HST} exposure time, thereby simultaneously including the majority of Virgo galaxies.
A 4-deg$^2$ deep stack from the first season of observation for our daily M87 Virgo field is shown in Figure~\ref{fig:virgo} along with the locations of six transients observed in the field (five are located in the background of Virgo and one is in the cluster).  These are YSE-observed SN~2020pf (a ZTF discovery but with simultaneous YSE observations), and YSE-discovered SNe~2019yub, 2020ndg, 2020lhy, 2020mvx, and finally AT~2020iuy, which is likely a stellar outburst with peak $M_{i} \approx -11.9$~mag.

\subsection{Criteria for Targeted Observations}

Although we wish to keep 50\% of our fields fixed for long-term monitoring, the other 50\% of YSE fields can be adjusted each year to follow new transients that meet our science objectives.  We hope to obtain targeted observations of $\sim$15 SNe per month.  This includes rare transients, young transients, or low-$z$ SNe~Ia to build an anchor sample for next-generation cosmology surveys.  Of these 50\% that can be adjusted, our survey requirements specify that 40\% of new fields should be within 15 degrees of existing fields to limit refocus overheads.  To move an existing YSE survey field to target a new SN, the survey field should meet these criteria (below, a setting field must have airmass greater than 1.5):

\begin{enumerate}
    \item To relocate an existing field to a new location, the existing field should either have a) no clearly rising transients and be within ten days of setting, or b) it should have no transients within either two weeks of maximum light or within a month of discovery. Certain rising transients that would have very poor light-curve coverage (e.g., if the field is near setting) would also be reasonable candidate fields to move.  Conversely, a current field with a post-max SN that is of high scientific value to the collaboration should not be discontinued unless it is near setting.
    \item If YSE science will substantially benefit from continued follow-up observations of the transients in the existing field, the field will not be moved.
    \item The location of the new field should be within 15 degrees of existing survey fields unless the target is of exceptionally high priority.
    \item In most cases, a new field should be three months or more from setting.
\end{enumerate}

\subsection{Interleaving YSE and ZTF Observations}

To maximize discoveries of young SNe and increase the effective cadence for many YSE and ZTF transients, we attempt to schedule YSE observations such that they precede ZTF observations by one calendar day.  Because our survey is $\sim$0.4--0.8~mag deeper than ZTF during dark time, we hope to discover fast-rising SNe or newly exploded SNe by observing a $>$1~mag rise between the Pan-STARRS observation and the ZTF observation approximately 21 hours later.  As ZTF reports transient discoveries with a median delay of just $\sim$10 minutes from the time the image is taken, our team is able to quickly combine the data streams to trigger rapid spectroscopic and multi-wavelength follow-up observations of fast-rising SNe.

In practice, predicting ZTF observations can be non-trivial, as ZTF uses a complex scheduling metric to avoid large coverage gaps rather than a fixed three-day cadence, and weather makes it difficult to anticipate whether an observation will be scheduled or carried out on a given night.  However, ZTF generally prioritizes fields by those with the longest time since they were last observed \citep{Bellm19}.  Over the first three months of YSE, during full nights of observing with clear weather at both Haleakala and Palomar, 61\% of our observations were taken a day before ZTF observations of the same field (compared to 33\% for uncoordinated observations; see Section~\ref{sec:yields} for examples of our data).  However, we find that during longer periods of good weather, we are more able to more successfully predict the ZTF observations.  Near the beginning of YSE, most fields were also concentrated within a narrow right ascension window, and these may be less likely to have ZTF observations evenly spread over three days. We note that ZTF will soon begin making their nightly observing plans public using the International  Virtual  Observatory  Alliance Observation Locator Table Access Protocol\footnote{\url{https://www.ivoa.net/documents/ObsLocTAP/index.html}.}, which will significantly improve our ability to plan overlapping observations.  Rubin will also announce their observing schedule via the same protocol, preparing us for potentially allowing YSE to supplement the Rubin cadence in 2023.

\subsection{Public Reporting}
\label{sec:reporting}

YSE reports all discoveries of likely transients, omitting likely variable stars or AGN, to the Transient Name Server\footnote{\url{https://wis-tns.weizmann.ac.il/}.} (TNS), the official International Astronomical Union mechanism for recording new transient discoveries.  Our discovery reports include the first epoch of photometry for each transient (we will provide the first two photometric epochs in the near future), and whenever a team member obtains a spectroscopic classification of an unclassified YSE transient, the spectrum and classification is sent to TNS as well.

Obtaining a second epoch before reporting a discovery is sometimes necessary for asteroid rejection; to avoid substantial $\sim$1-minute refocusing overheads for Pan-STARRS when slewing more than 15 degrees, our observations in different filters are spaced only $\sim$4 minutes apart.  This makes asteroid rejection more difficult, occasionally necessitating a second observation or confirmation from a second survey in order to confirm that a YSE alert is a bona fide transient.  When YSE observations for a new field begin, AGN contamination is also significant; as our team has gained vetting experience we have become more conservative in promoting candidates during the first epochs of a given field to avoid reporting AGN.  In most cases, however, the colors and presence of nearby host galaxies make transient discoveries unambiguous.  In cases where it is unclear if a transient is a true SN versus an asteroid, variable star, or image artifact, we wait to send these events to TNS until additional epochs of data can clarify whether or not they are bona fide transient phenomena.

\subsection{Anticipated SN Yields}
\label{sec:sim_yields}

\begin{figure}[ht]
    \centering
    \includegraphics[width=3.4in]{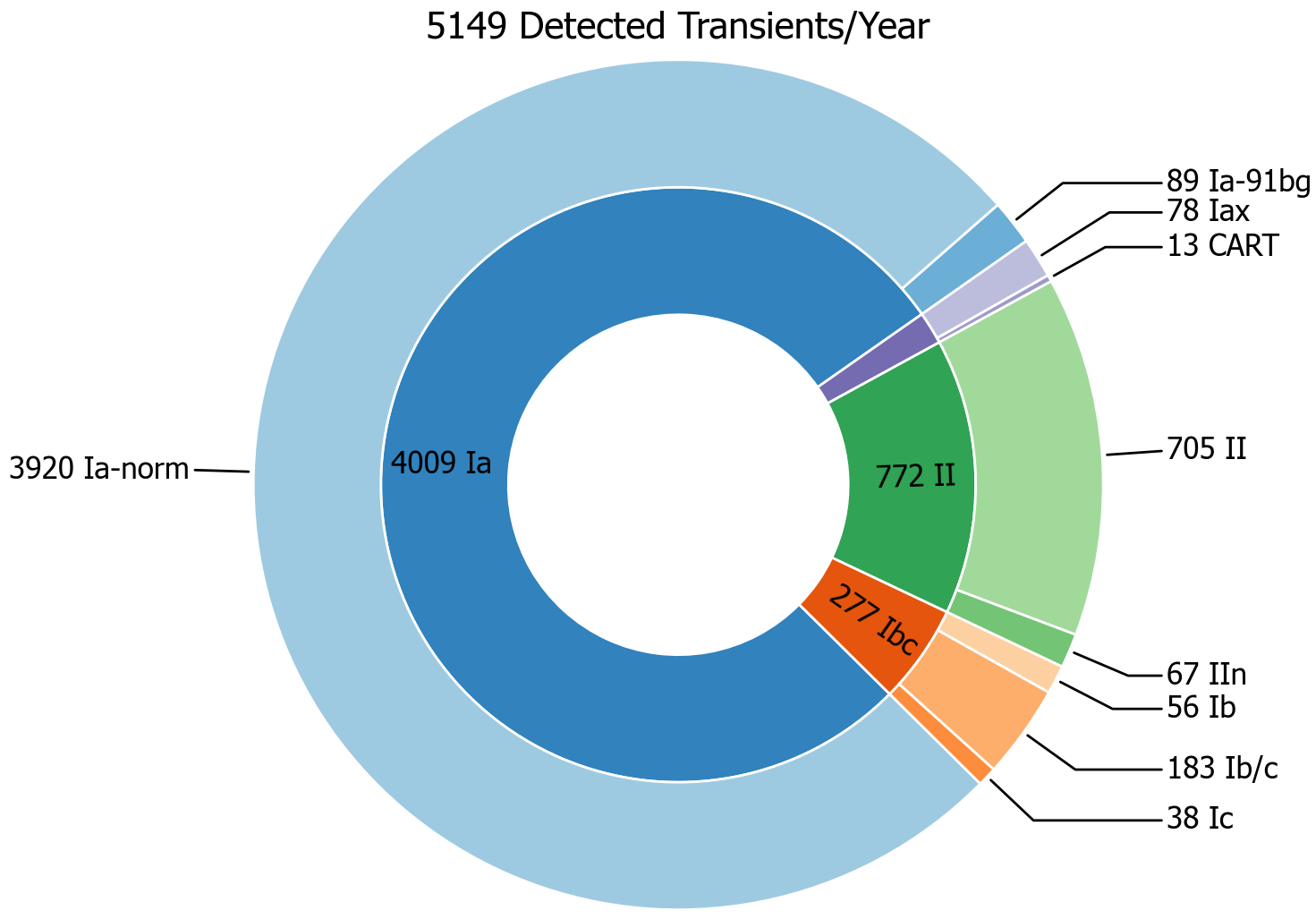}
    \caption{Predicted number of SNe per year with at least three YSE detections of ${\rm S/N} > 5$, as determined from SNANA simulations.  SN models are from PLAsTiCC (\citealp{Kessler19} and references therein). We predict 3920 SNe~Ia, 277 SNe~Ib/c, 705 SNe~II, and 67 SNe~IIn each year of full survey operations.  We note that PLAsTiCC templates for certain models, including a subset of SNe~II and SNe~Ib/c, were not given subtypes and thus do not have subtypes here.}
    \label{fig:surveysims}
\end{figure}

To evaluate the effectiveness and expected SN discoveries from the survey strategy presented above, we simulated  the YSE survey using the SNANA software \citep{Kessler10}.  We include models of different SN types and subtypes from the PLAsTiCC SN classification challenge \citep{Kessler19}, including SN\,Ia, SN\,Iax, SN\,II, SN\,Ib/c, ILOT, and CART.  We omit TDEs, SLSN-I and kilonovae due to large uncertainties on their rates, as well as purely theoretical models and non-transient models (e.g., variable stars and AGN).  We use anticipated survey depths from \citet{Chambers16}, moon phase-dependent sky noise to match 3$\pi$ observations, SN rates from \citet{Strolger15} and \citet{Dilday08}, normalization of those rates at $z = 0$ from PLAsTiCC (and references therein; most rates originate from \citealp{Li11}), and a nominal ZTF survey to match their reported depths with ZTF observations scheduled 21 hours after YSE.  Details of the simulation methodology and the effect of choosing alternate survey strategies are given in Appendix~\ref{sec:sim_appendix}.

\begin{figure}[ht]
    \centering
    \includegraphics[width=3.4in]{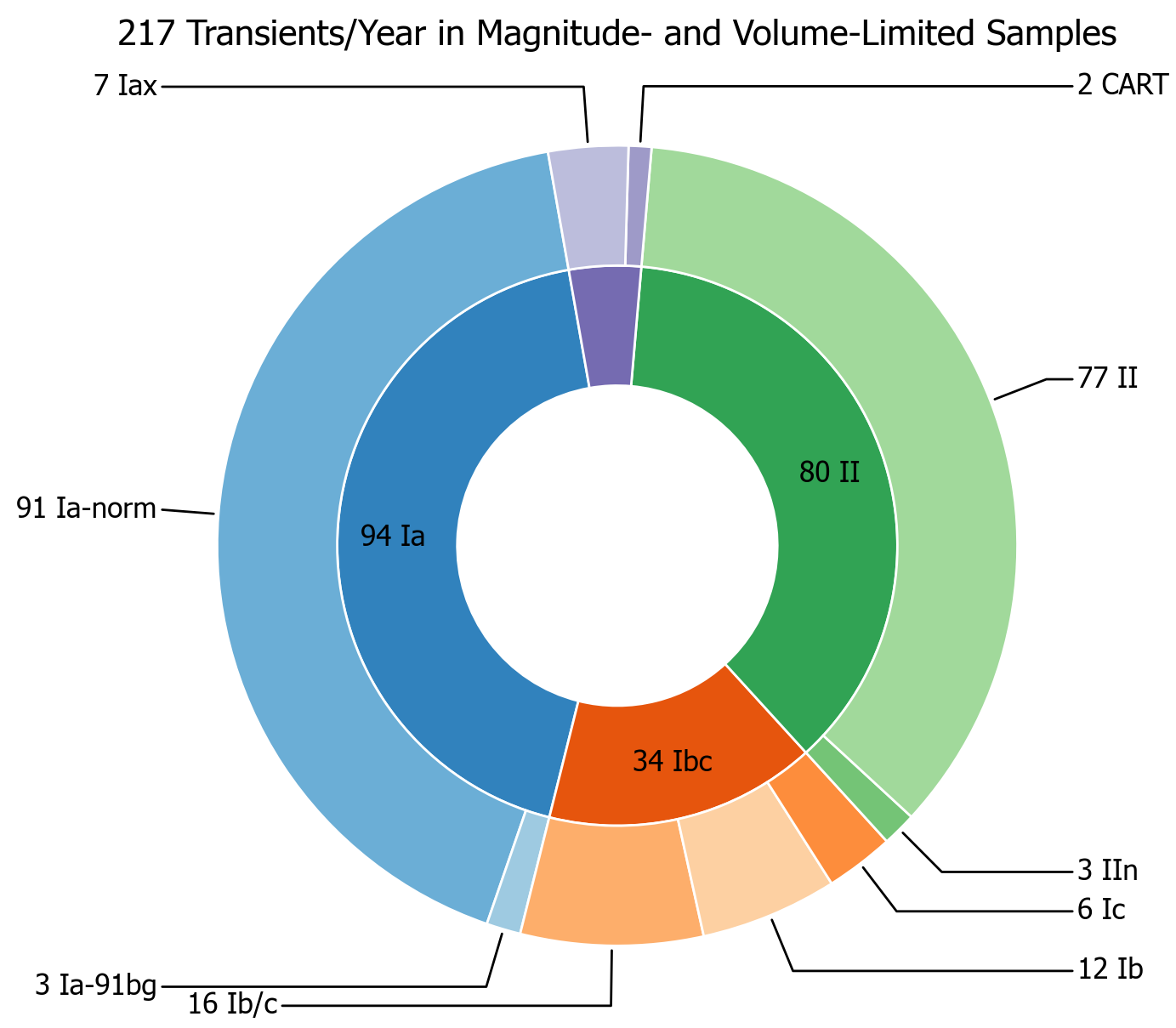}
    \caption{Same as Figure~\ref{fig:surveysims}, but for YSE SNe per year meeting the criteria of our volume- or magnitude-limited surveys, $<$250~Mpc or $r < 18.5$~mag (a magnitude at which we expect to be able to obtain spectroscopic classifications).  We predict 94 SNe~Ia, 34 SNe~Ib/c, 77 SNe~II, and 3 SNe~IIn per year.  SN models are from PLAsTiCC (\citealp{Kessler19} and references therein).  We note that PLAsTiCC templates for certain models, including a subset of SN~II and Ib/c, were not given subtypes.  Sixty-one transients (36\% of the volume-limited sample) are both more distant than 200~Mpc and fainter than $\sim$18.5, the limits of the ZTF ``Census of the Local Universe" and magnitude-limited samples.}
    \label{fig:surveysims_magvol}
\end{figure}

Summary statistics for the nominal simulated YSE survey design are shown in Figure~\ref{fig:surveysims}.  Due to its depth --- YSE is able to find SNe\,Ia at redshifts up to nearly $\sim$0.3 --- simulations predict 5149 SNe per year having at least three detections with a signal-to-noise ratio (S/N) $>$ 5, including 3920 SNe~Ia, 277 SNe~Ib/c, 705 SNe~II, and 67 SNe~IIn.  Our magnitude- and volume- limited samples, comprising SNe with $m < 18.5$~mag and $D < 250$~Mpc, respectively, are expected to contain a total of 217 SNe per year (Figure~\ref{fig:surveysims_magvol}; 49 SNe are members of only the magnitude-limited survey, 109 are members of only the volume-limited survey, and 59 are members of both).

In Appendix~\ref{sec:sim_appendix}, we discuss the results of several other survey designs explored by our team, including variations in exposure time, three or four filters per day, and a survey design focused on either blue or red filters.  We caution that some simulated yields are affected by small-number statistics.  As exposure time increases, the number of discovered SNe tends to increase, but begins to flatten after 25 seconds, as would be expected from Figure~\ref{fig:depth}.  Surveys consisting of three or four filters per day would be excellent for some science cases, but would drastically reduce the discovery space for transients, resulting in a 40--50\% reduction in the number of transient discoveries.  A blue-focused survey was primarily ruled out because it offers a less unique discovery space compared to existing time-domain surveys, while a survey focusing only on $riz$ observations would be expected to reduce young SN yields as most young SNe are expected to be blue.  Our nominal survey design $-$ 27s exposures with a $gr$, $gi$ observing sequence during dark time $-$ was settled on as the best choice for YSE, especially given the characteristics of the Pan-STARRS system and our allocated 7\% fraction of time on each telescope.

\begin{figure}
    \centering
    \includegraphics[width=3.4in]{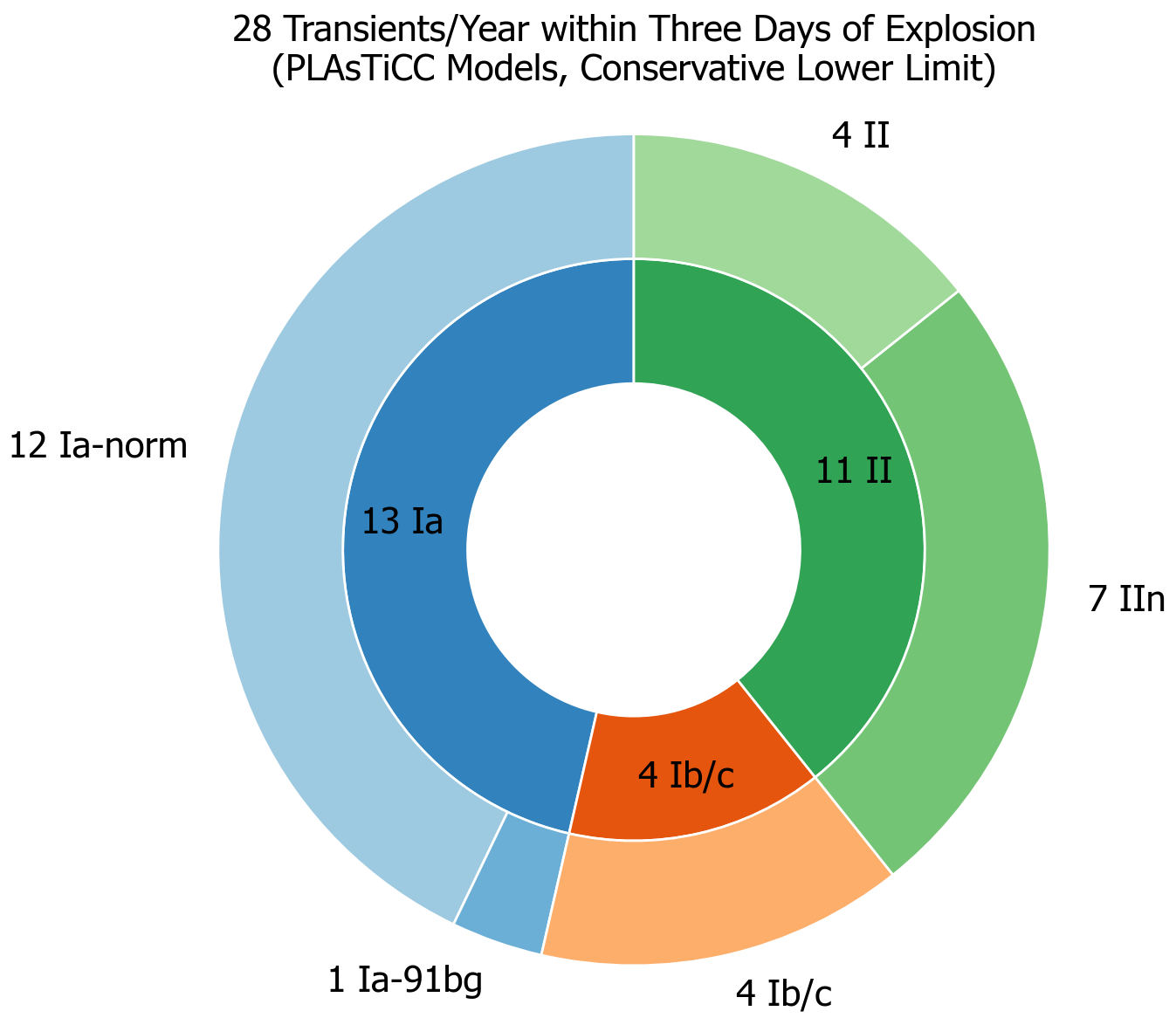}
    \caption{Same as Figure~\ref{fig:surveysims}, but for YSE SNe per year discovered within three days of explosion.  Simulations yield 13 SN\,Ia, 3 SN\,Ib/c, and 11 SN\,II.  We expect the simulation to underestimate to true number of young SNe since the SALT2 model for SNe~Ia has a slower early rise than, for example, the SNe discovered by \citet{Jiang20}, and few PLAsTiCC CC~SN models include a shock breakout cooling stage.}
    \label{fig:youngsn}
\end{figure}

Finally, Figure~\ref{fig:youngsn} and Table~\ref{table:yse_strategy} show the anticipated yields of SNe less than three days old assuming that PLAsTiCC models are correct realizations of young SN brightness and color.  For example, PLAsTiCC simulations of SNe~Ia are based on the SALT2 model, which has a slower rise at early times than observations of SNe~2017cbv, 2018oh, and a number of SNe discovered by \citet{Jiang20}.  Since SNe~IIb often show luminous shock breakout cooling \citep[e.g.,][]{Richmond94,Arcavi11,Kilpatrick17,Fremling19}, they are often easy to detect early as well, yet there are no SN~IIb templates in the PLAsTiCC set.  It is therefore difficult to truly predict the number of young SNe that will be discovered by YSE, but we note that our team has already observed a handful of young SN candidates, including SNe~Ia 2020pf, 2020fci, 2020ioz, 2020juq, 2020nbo and CC~SNe 2020pni, 2020kpz, and 2020tlf that are almost certainly within three days of explosion or less (see Section~\ref{sec:yields}).

\begin{figure}
    \centering
    \includegraphics[width=3.4in]{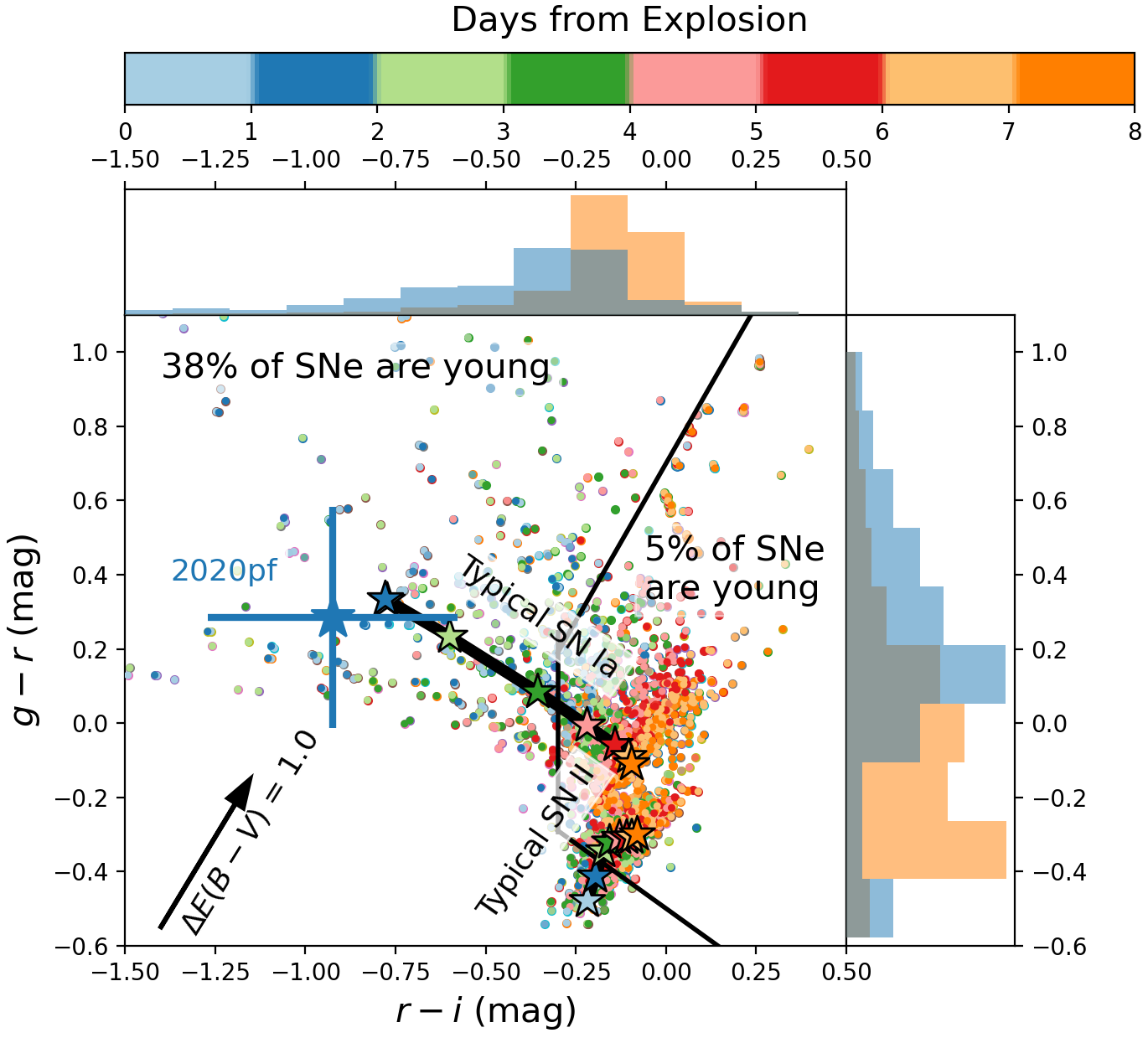}
    \caption{Color-color diagram for ``noise-free'' YSE SNe as simulated by SNANA with colored points representing the time since explosion as described by the above color bar.  The black arrow indicates the direction and magnitude an object would experience if its dust reddening were increased by 1~mag with $R_{V} = 3.1$.  The black line defines a relatively simple cut that adequately identifies SNe within three days of explosion, with 38\% of all objects to the left of the line being young, while only 5\% of those to the right of the line are young.  We highlight the path a typical SN~Ia and SN~II (star symbols) travel as they evolve from explosion.  We note that the $g-r$ and $r-i$ colors are particularly useful for selecting young SNe~II and SNe~Ia from the entire population, respectively.  SN~2020pf at an epoch of $\sim$2~days after explosion from YSE and ZTF data is displayed as a star with error bars; it is consistent with the simulations and in the defined ``early'' region.  To the top and right are histograms of the $g-r$ and $r-i$ colors for SNe before $+$3~days from explosion (blue) and after $+3$ days from explosion (orange).}
    \label{fig:youngsn_colorcut}
\end{figure}

Figure~\ref{fig:youngsn_colorcut} also shows the simulated colors of young SNe; we find that a simple color cut could identify a subset of young SNe with $\sim$40\% accuracy, which is a reasonably high confidence given the follow-up resources available.  We find that $g-r$ colors are useful for separating young versus old CC\,SNe and YSE $r-i$ colors appear to be excellent indicators of young SNe\,Ia, complementing $g-r$ observations from YSE and ZTF.  Our team is working on more sophisticated classification methods using existing observations of young SNe.

While these simulations are state of the art for transient survey planning, they still lack several details for precise estimates. Although early YSE survey yields are returning the expected number of volume- and magnitude-limited SNe, just 26\% of transients to date are above the median predicted survey redshift of $0.19$.  This is likely because our algorithm for measuring YSE photometry has been obtaining inflated uncertainties, effectively reducing the S/N of YSE photometry and our detection efficiency.  We identified this issue by implementing a second photometric reduction algorithm and are currently working to implement a fix to our pipeline photometry.  Other contributing factors may include biases in the photo-$z$ determinations, as reliable redshifts are more difficult to estimate for faint or undetected host galaxies.  Additional stacked observations may also reduce the noise in the template images, which simulations assume to be negligible.  Finally, efficiency losses in the machine-learning ``real/bogus" algorithm (Section~\ref{sec:broker}) may also reduce the number of discovered SNe as a function of S/N.  Nevertheless, we have observed or discovered \ntotalsne\ transients over the first ten months of the survey while observing at half of our nominal area and experiencing $\sim$2~months of down time.

\subsection{Impact on YSE Science Drivers}
\label{sec:science_impact}

Given the YSE survey design and simulations described above, we briefly summarize a number of ways in which the YSE data will address the science drivers identified in Section~\ref{sec:science}:
\begin{itemize}
    \item {\bf Young SNe}.  YSE will discover at least 2 SNe within three days of explosion per month, with a larger sample likely due to known deficiencies in the PLAsTiCC models used to generate YSE simulations.  Many of these SNe will have $i$ or $z$ photometry at very early times and can be discovered by YSE $\sim$1 day before they are found by other surveys.  Several examples from early YSE data are discussed in Section~\ref{sec:yields}.
    \item {\bf Red and Rare Transients}.  The $iz$ coverage of YSE will aid the identification of very red transients.  YSE's additional $\sim$1.2-mag $i$-band depth compared to ZTF will probe a unique volume equal to $\sim$76-90\% of the ZTF $i$-band survey volume (Table \ref{table:yse_comp}).  YSE is also the only time-domain survey currently observing in the $z$ band.  Within our volume-limited sample out to 250~Mpc, we will discover transients down to $M_i \approx -15.5$~mag and $M_z \approx -16.5$~mag.  We also expect YSE to observe two glSNe per year in regular sky scans and up to five glSNe per year in stacked images based on projections from \citet{Wojtak2019}.  YSE fields to date include $\sim$1400 galaxies near enough for YSE to discover luminous red novae, though their volumetric rate is not well known \citep{Pastorello19}.
    \item {\bf Cosmology}.  The YSE SN~Ia sample will be a mix of targeted SN~Ia observations to increase the $z \lesssim 0.05$ sample size, using a strategy modeled after the Foundation Supernova Survey \citep{Foley18}, and untargeted observations, which will have a median redshift of $z \approx 0.12$ (Appendix~\ref{sec:sim_appendix}).  The untargeted sample will include up to 350 $z < 0.1$ SNe per year with a well-understood, magnitude-limited selection and lower uncertainties from peculiar velocities than other low-$z$ samples.  The large sample with $z > 0.1$ will serve as an important sample for measuring $w$, testing general relativity, and helping to understand CC\,SN contamination in future cosmology analyses.  The photometric calibration of the sample will be more precise than any non-Pan-STARRS low-$z$ sample.
    \item {\bf Black Holes and Tidal Disruption Events}.  We did not include TDEs in survey simulations as their redshift-dependent rates are uncertain, as is the efficiency of detecting them in the bright cores of galaxies; based on the survey magnitude limit alone with the rates and luminosities from \citet{vanVelzen20}, and including an approximate mean $M_g = -17.5$~mag, we would expect $\sim$20 per year --- and YSE has already observed two TDEs in untargeted observations just in early survey data (a third, AT~2020nov, was targeted; Section~\ref{sec:yields}) --- but due to the difficulty of detecting transients in the cores of galaxies, it is unclear if the expected number from these order-of-magnitude estimates is  overly optimistic.  YSE survey data will also contain tens of thousands of AGN for variability studies and its over-sampled PSF may make it easier to detect transients at the bright cores of galaxies compared to other low-$z$ surveys.
    \item {\bf Preparation for the Rubin Observatory}.  YSE will measure $griz$ light curves for hundreds of spectroscopically classified transients and thousands of transients in total, including hundreds of detections within the first few days of explosion.  This will allow YSE to serve as a unique and invaluable training sample for the Rubin Observatory, adding fidelity to early time and full light curve classifications and constraining the rates of rare classes that will exist in Rubin survey data.
    \item {\bf Magnitude- and Volume-limited Census of SNe}.  The $iz$ coverage of YSE will improve the census of red transients and the YSE redshift range will allow a longer lever arm on the redshift-dependent rates of many transients compared to other ongoing surveys.  We plan to spectroscopically classify every SN that is either brighter than $r \approx 18.5$~mag at peak or nearer than 250~Mpc, a sample of $\approx$217~SNe per year (Section~\ref{sec:sim_yields}) to constrain both the rates and luminosity functions of SN types and subtypes.  Approximately 61 of these SNe (36\% of the volume-limited sample and 28\% of the total) will be both fainter than $r \approx 18.5$ and more distant than 200~Mpc, the limits of the ZTF magnitude-limited and ``Census of the Local Universe" (CLU) samples.  Additionally, we include photometric redshift estimates in building our volume-limited sample in order to avoid completeness limitations that affect spectroscopic catalogs of the local volume (and the CLU sample to some degree).
\end{itemize}

\begin{table}[]
\caption{\fontsize{9}{11}\selectfont Stacked Image Depths}
    \centering
    \begin{tabular}{lrrrrr}
      \hline \hline\\[-1.5ex]
      && \multicolumn{4}{c}{Depth}\\
      Survey & Area & $g$ & $r$ & $i$ & $z$\\
      & (deg$^{2}$) & \multicolumn{4}{c}{(mag)}\\
    \hline\\[-1.5ex]
PS1 3$\pi$                  & 30000 & 23.3 & 23.2 & 23.1 & 22.3\\
SDSS                        &       15044 & 23.3 & 23.1 & 22.3 & 20.8\\
DES DR1$^{\rm a}$           &  5186 & 24.3 & 24.1 & 23.4 & 22.7\\
KiDS DR4                    &        1006 & 25.1 & 25.0 & 23.7 & \\
\hline\\[-1.5ex]
YSE$+$3$\pi$ 1-yr                  &  $\sim$5000 & 23.6 & 23.7 & 23.6 & 22.8\\
YSE$+$3$\pi$ 3-yr (deep$^{\rm b}$) &  $\sim$2000 & 24.0 & 24.2 & 24.1 & 23.3\\
YSE$+$3$\pi$ 3-yr (wide$^{\rm c}$) &  $\sim$7000 & 23.6 & 23.7 & 23.6 & 22.8\\
\hline\\[-1.5ex]
    \end{tabular}
    
    \begin{minipage}{8.5cm}
    $^{\mathrm{a}}$ 10-$\sigma$ limits.\\
    $^{\mathrm{b}}$ Corresponding to the region monitored for the entire 3 years.\\
    $^{\mathrm{c}}$ Corresponding to the region monitored for 1--2 years.

    {\bf Note.} YSE projected stacked image depths compared to several large-area surveys with observations in the $griz$ bands.  Limits for SDSS, PS1 3$\pi$, DES, and the Kilo-Degree Survey (KiDS) are from \citet{York00}, \citet{Chambers16}, \citet{Abbott18}, and \citet{Kuijken19}, respectively.
    \end{minipage}
    \label{table:stacks}
\end{table}

In addition to addressing these time-domain science questions, YSE will create one-year stacked images over approximately 7000 deg$^2$ of sky and three-year stacked images over an additional $\sim$2000 deg$^2$ of sky.  Using the typical read noise and moon-dependent sky noise values from \citet{Chambers16} and assuming 30\% loss due to weather and 24\% due to pixel masking, we roughly estimate that in one-year stacks, YSE$+$3$\pi$ will reach depths of $griz \approx 23.6, 23.7, 23.6, 22.8$~mag.  In the three-year stacks, YSE$+$3$\pi$ will reach depths of $griz \approx 24.0, 24.2, 24.1, 23.3$~mag.  Comparisons to the depths of several other large-area surveys are shown in Table \ref{table:stacks}.

\begin{figure*}
    \centering
    \includegraphics[width=5.2in]{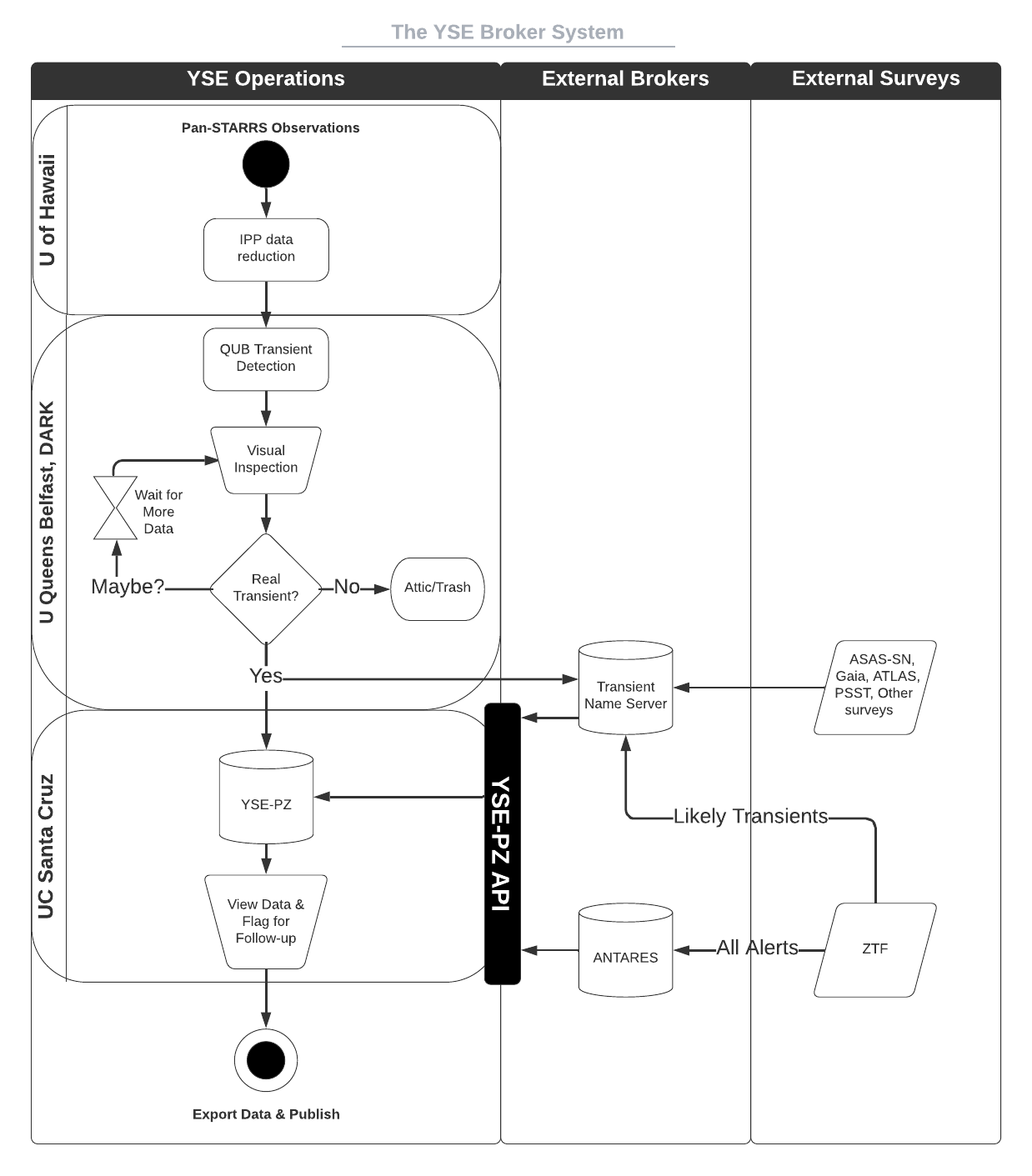}
    \caption{Activity diagram for the YSE Transient Broker process as described in Section~\ref{sec:broker}.  The left column indicates the progress and locations of the data and metadata.  Data are reduced in Hawaii and sent to the Transient Science Server at Queens University Belfast; members of our team at DARK (U.\ Copenhagen) visually inspect each event that passes machine learning cuts.  Real transients are reported to TNS, likely AGN or asteroids are given ``attic'' designations, and image artifacts are sent to ``garbage.''  Finally, YSE data are combined with public data from other surveys at which point our team sorts the discoveries and requests follow-up observations of interesting transients.  In addition to the workflow shown here, YSE-PZ can request forced photometry upper limits directly from the IPP for new ANTARES or TNS transients without Pan-STARRS detections.}
    \label{fig:yse_schematic}
\end{figure*}

\begin{figure*}
    \centering
    \includegraphics[width=6in]{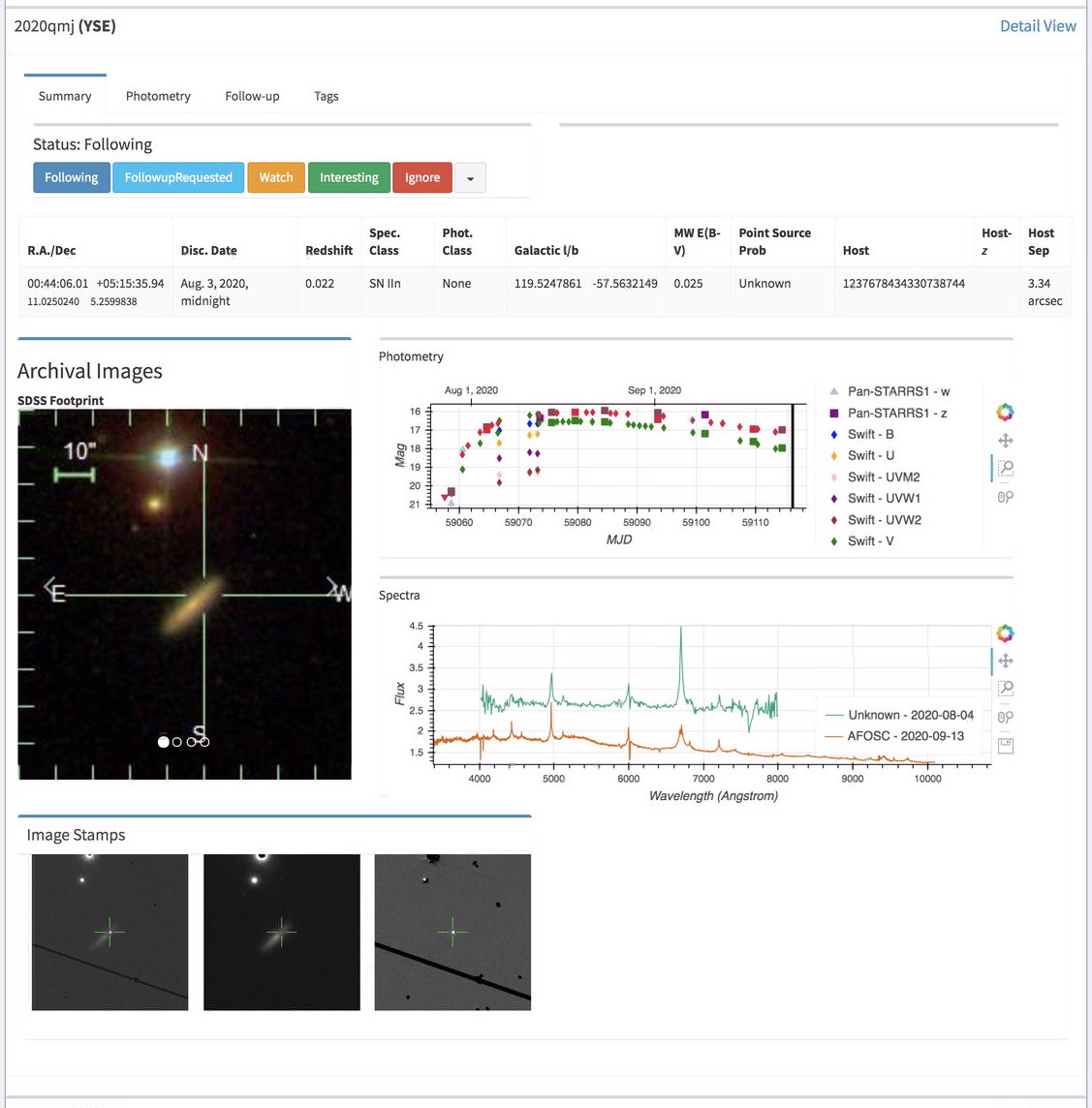}
    \caption{Example ``transient summary'' page from the YSE-PZ web application. Transient summary pages show YSE image stamps, spectra, photometry, archival images, and metadata, with a more extensive set of data and metadata available via the ``detail view.''}
    \label{fig:yse_pz}
\end{figure*}

\section{Vetting and Following Transients}
\label{sec:broker}

Efficiently working with large time-domain data sets to identify transients of interest for follow-up observations and facilitating large statistical analyses with the data are key challenges of transient discovery searches \citep{Bloom12,Kasliwal19,vanderWalt19}.  Efficiently separating data artifacts from bona fide transients \citep[e.g.,][]{Duev19} is also critical.  YSE in particular requires a robust and efficient broker process for discovering and following transients.  

Once YSE data are taken, they are first processed by the IPP difference image pipeline and sent to the Transient Science Server at Queens University Belfast \citep{Smith20}.  The Transient Science Server then determines the likelihood that each SN is a bona fide transient, including a visual inspection stage performed by our team.  From there, possible transients are sent to a transient database and web application housed at UC Santa Cruz, which we refer to as ``YSE-PZ.''  The full YSE broker process is summarized in Figure~\ref{fig:yse_schematic} and the details of this process are discussed below.

\subsection{The Transient Science Server}
\label{sec:tss}

The Transient Science Server removes YSE detections from known AGN, variable stars, and moving objects using a library of catalogs (see \citealp{Smith20} for details) and employs both pixel-based and catalog-based machine learning algorithms --- i.e., the algorithm is based on both the pixels themselves and on photometric metadata such as PSF shape $-$ to remove likely image artifacts, and assign a ``real-bogus'' score to each detection.  Candidates are included on the by-eye vetting list if the real-bogus score considers them to be real with at least 32\% confidence; we choose this threshold because the training procedure estimates it will yield a high fraction of real candidates (75\%), while rejecting just 1\% of bona fide transients.  We require 3.5-$\sigma$ detections in at least two separate images (possibly over multiple epochs) to add a source to the list of candidates to be vetted.  The machine-learning algorithm is trained using slow-moving asteroids as the ``good'' objects; we plan to retrain using real transients in more realistic host-galaxy environments in the future.  Additional details of the algorithm are given in \citet{Smith20}.

If members of our team determine that a candidate is likely a bona fide transient, they are immediately sent to the Transient Name Server along with the first epoch of photometric data.  If we are unsure, the transients remain on a ``possible'' list until more data have been collected.  Both the transients on the ``good'' and ``possible'' lists are ingested to YSE-PZ for combining with data from other surveys and making follow-up decisions.

\subsection{YSE-PZ: The YSE Transient Management System}

The YSE-PZ application (YSE-PrioritiZe) is designed to ingest every transient reported to the Transient Name Server\footnote{\url{https://wis-tns.weizmann.ac.il/}} to facilitate queries of the data, combine YSE transients with external data, store follow-up data obtained by our team, and enable easy communication among team members for vetting and following new transients.  YSE-PZ is a Django-based application built on a MySQL database.  The code base is publicly available at \url{https://github.com/davecoulter/YSE_PZ} and we encourage collaboration and new contributors.  An example of the web application interface is shown in Figure~\ref{fig:yse_pz}.

YSE-PZ will be described in detail in Coulter et~al.\ (in prep.).  Here, we summarize key features that allow our team to prioritize and organize follow-up observations of transients:

\begin{figure*}
    \centering
    \includegraphics[width=7in]{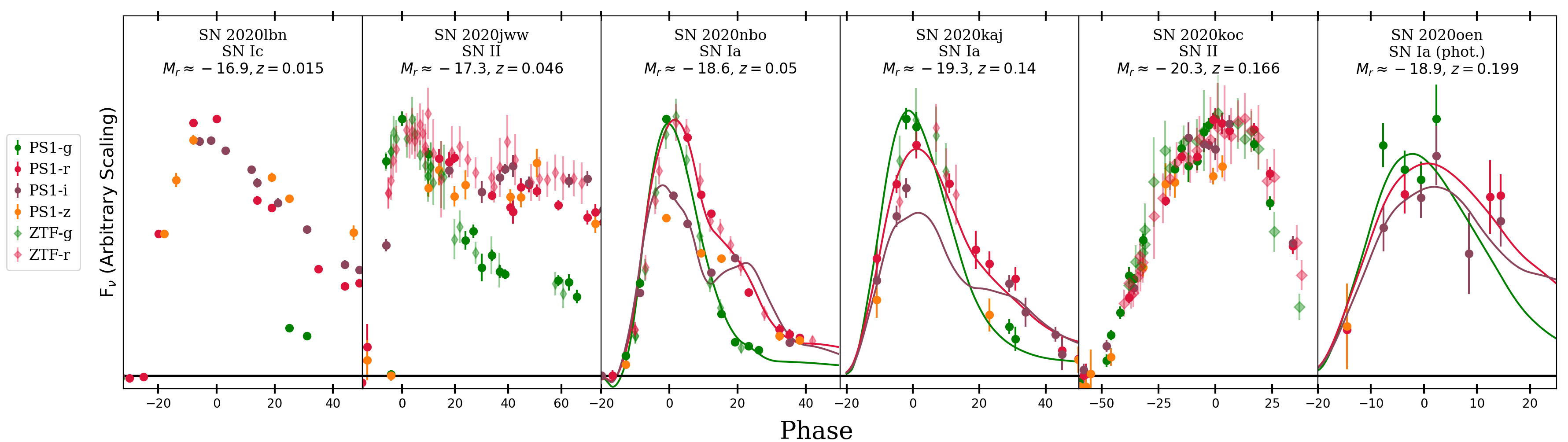}
    \caption{Examples of YSE SN light curves by increasing redshift, from $z = 0.02$ to $z = 0.2$.  For the SNe~Ia, we also display the best-fit SALT2.4 template light curves \citep{Guy10,Betoule14} for illustration.  ZTF $gr$ light curves are displayed as lightly shaded diamonds when available, which effectively doubles the cadence of nearby SNe in our sample (though low-$z$ SN~Ic 2020lbn, left, has no public ZTF data).  We note that CC~SNe~2020hrw and 2020koc are somewhat unusual SNe, as SN~2020hrw is likely a SN~1987A-like SN and SN~2020koc is unusually luminous for a SN~II (although perhaps not quite luminous enough to be considered a SLSN).}
    \label{fig:real_lcs}
\end{figure*}

\begin{enumerate}
    \item {\bf Data ingestion from ZTF and TNS}.  We ingest ZTF alerts using the ANTARES broker \citep{Narayan18} for events in our survey fields that have not yet been reported to TNS.  We also ingest every transient reported to TNS using the TNS API.  For transients that have been reported to TNS, we ingest ZTF photometry from MARS\footnote{\url{http://mars.lco.global/}.}.
    \item {\bf Dashboard and Initial Transient Vetting}.  Transients are ingested into YSE-PZ with a status of ``New'' and then parsed into categories of ``Watch,'' ``Interesting,'' ``FollowupRequested,'' ``Following,'' or ``Ignore'' based on characteristics such as brightness, redshift, classification, or color.  Transients that meet the YSE volume- or magnitude-limited survey criteria are prioritized for followup by moving them to ``Interesting'' or ``FollowupRequested'' and subsequent queries are used to flag transients that brighten to the point where they are included in the magnitude-limited criteria.  YSE-PZ compiles a large amount of data and metadata to aid in the vetting process, including survey images, archival images, other external data (e.g., {\it HST} or {\it Chandra} imaging), Milky Way extinction, photometry, spectra, classifications, and discovery details.  Our team sorts through new transient discoveries daily.
    \item {\bf Follow-up Requests}.  YSE-PZ allows follow-up observations to be requested on any telescope for which our team has access.  Follow-up requests are added to a custom page for a given observing night and resource, where observers can prioritize the requests and schedule observations. Finder charts using Pan-STARRS 3$\pi$ imaging may be generated through the web interface and we plan to add additional tools to facilitate scheduling in the near future.
    \item {\bf Queries}.  YSE-PZ allows users to query on any field in the database.  A query, once completed and saved, can be added to a user's personal ``dashboard'' to allow each user to be alerted to transients that are relevant to their science interests.  Example queries include the YSE volume-limited sample of transients within 250~Mpc, or transients rising by more than 0.5~mag in a day in a given filter.  As new transients are ingested, users can choose to be alerted via email or text message to transients that match a given query to enable extremely rapid follow-up observations.
    
\end{enumerate}

YSE-PZ is under active development, and new features are continually being added.  A well-developed, open-source, robust transient management framework is vital to our science goals and a key part of developing tools for the next decade of transient science.

\begin{figure}
    \centering
    \includegraphics[width=3.4in]{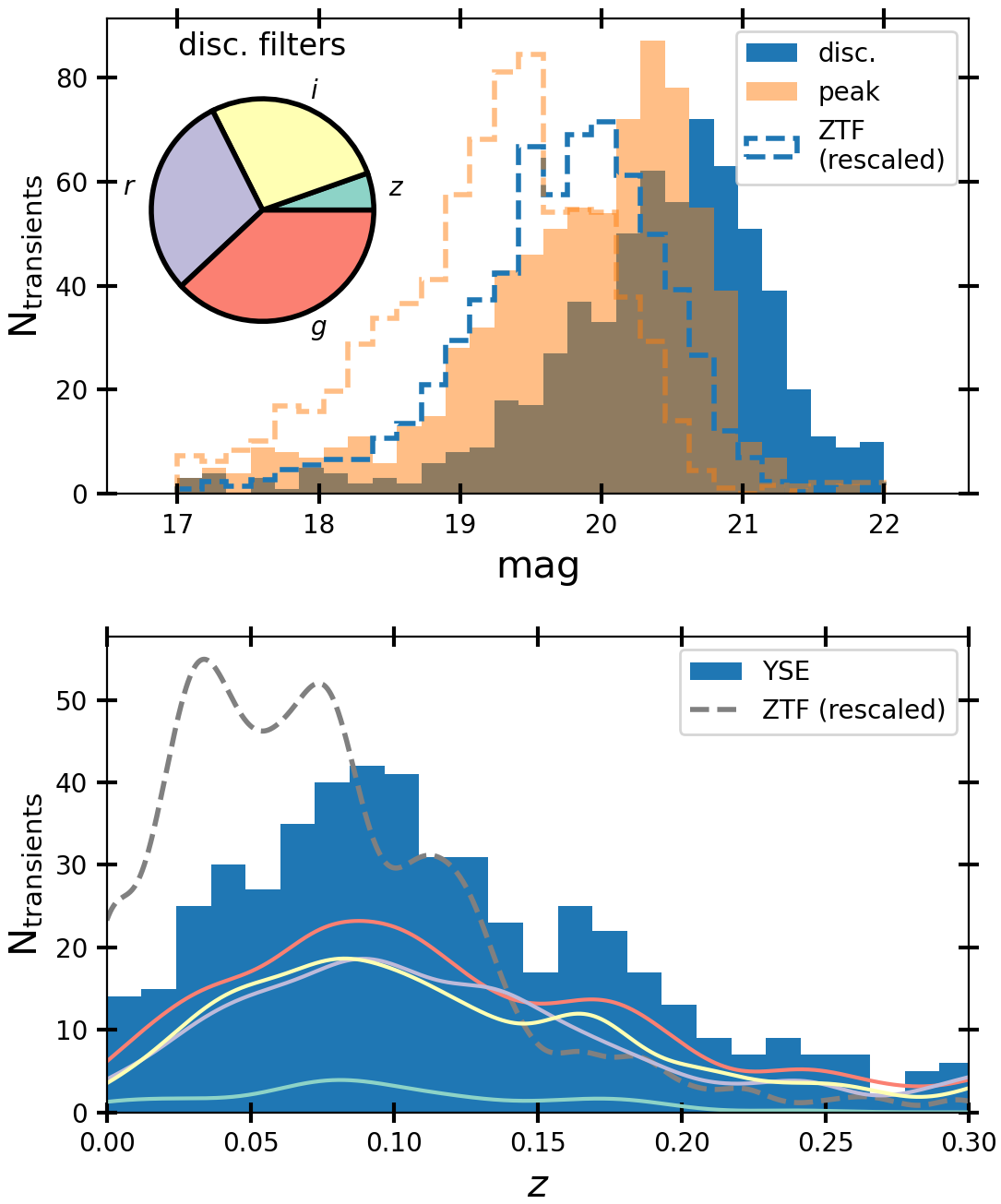}
    \caption{Top: Approximate peak and discovery magnitudes for YSE transients to date, including targeted follow-up observations, with the filter combinations used at each discovery epoch or first observation shown in the pie chart.  Histograms of 2020 ZTF discoveries, re-scaled to match the total number of YSE discoveries, are shown with dashed lines.  Estimated peak magnitudes are taken from the brightest epoch, regardless of filter, and include publicly reported observations when available.  Discovery magnitudes are more often from the bluer of the two filters at the discovery epoch, as the bluer observation is typically performed first.  Bottom: redshift distribution (using an internal photo-$z$ estimate trained on SDSS data, when necessary) for the YSE sample to date.  The normalized ZTF redshift histogram for 2020 discoveries to date is shown in grey.  Lines show Gaussian kernel density estimates for discoveries in each filter with colors corresponding to the pie chart in the top plot.  Photo-$z$ outliers may artificially increase the numbers in some high-redshift bins.}
    \label{fig:inithists}
\end{figure}

\section{Survey Status and Discoveries to Date}
\label{sec:yields}

\begin{table}
\caption{\fontsize{9}{11}\selectfont Initial Statistics of SNe Observed by YSE}
  \centering
\begin{tabular}{lr}
  \hline \hline\\[-1.5ex]
&N$_{\mathrm{transients}}$\\
  \hline\\[-1.5ex]

All SNe$^{\rm a}$&916\\
Discovered by YSE$^{\rm a}$&363\\
Spec.-confirmed SNe&\nspecclass\\
$r \lesssim 18.5$&125\\
$D \lesssim 250$ Mpc&180\\
SNe~Ia with ${\rm phase} < -10$~days&30\\

\hline\\[-1.5ex]
\multicolumn{2}{l}{
\begin{minipage}{6cm}
$^{\rm a}$ Excluding candidate transients deemed later to be likely variable sources such as AGN.

{\bf Note.} Transient discovery statistics between Nov. 24th, 2019 and Oct. 1$^{\rm st}$, 2020.  We have been at 50\% of our full observing allocation since early January, but lost $\sim$2 months due to weather and telescope malfunctions.
\end{minipage}}
\end{tabular}
\label{table:yse_stats}
\end{table}

\begin{figure}
    \centering
    \includegraphics[width=3.4in]{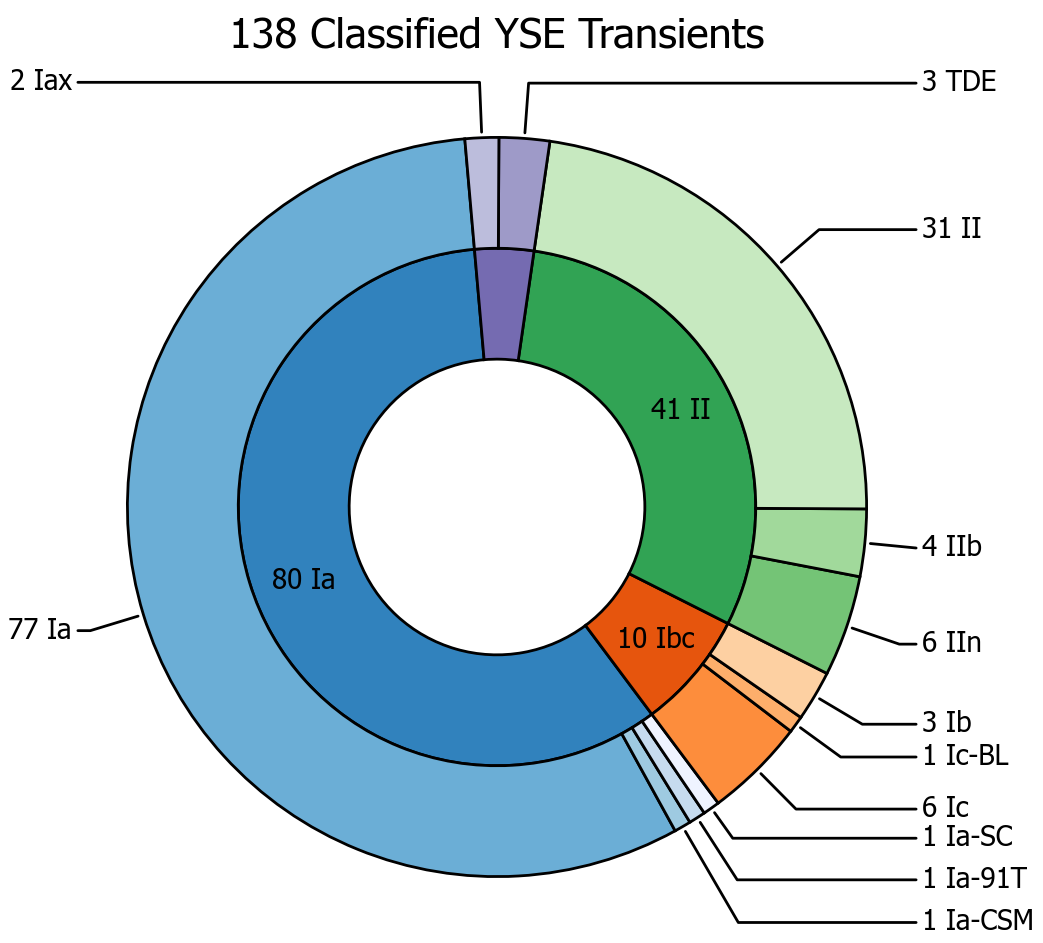}
    \caption{To-date, by-type breakdown of the \nspecclass\ spectroscopically classified YSE SNe prior to Oct.\ 1$^{\rm st}$, 2020.}
    \label{fig:class_pie}
\end{figure}

\begin{figure*}
    \centering
    \includegraphics[width=7in]{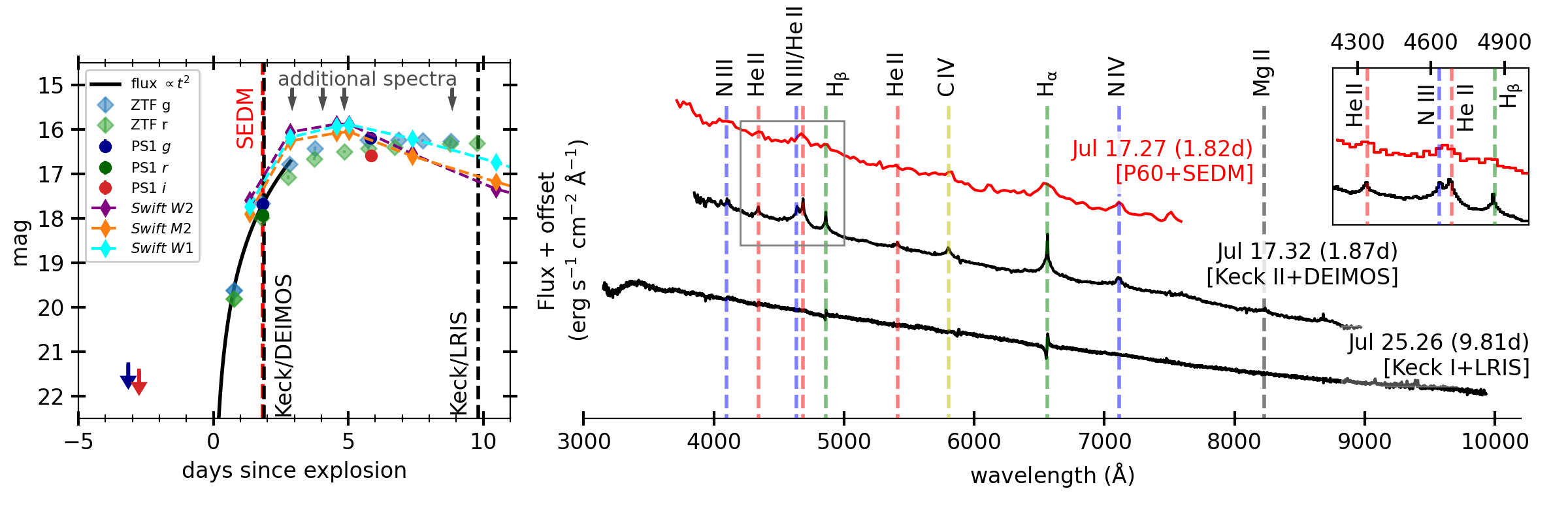}
    \caption{Early light curve (left) and spectra (right) for CC~SN 2020pni.  ZTF detected SN~2020pni approximately 0.8~days after explosion \citep{Forster20_20pni}, with deep YSE non-detections (arrows, data taken $\sim$4~min.\ apart but are horizontally spaced for visual clarity) coming two days prior.  {\it Swift} follow-up observations were triggered, and Palomar 60-inch \citep{Bruch20_20pni,Bruch20b_20pni} and Keck spectra showing the flash-ionized CSM were taken all within two days of explosion.  Flash ionization features had disappeared by approximately five days after explosion.  Our full analysis will be presented in Terreran et~al.\ (in prep.).}
    \label{fig:20pni}
\end{figure*}

YSE has discovered or observed a total or \percentdisc\% of IAU-reported 2020 transients within our survey fields.  YSE has announced a total of \ndiscreports\ discoveries as of October 1$^{\rm st}$, 2020.  
Though early discovery announcements included significant AGN contamination, our team has become much more accurate over time at removing AGN, asteroids, and artifacts (burns, persistence, etc.) that passed initial cuts; after inspecting the full YSE light curves to remove these objects, we estimate that YSE discovered or observed \ntotalsne\ bona fide transients prior to October 1$^{\rm st}$, 2020.
A summary of our discovery and follow-up statistics are reported in Table~\ref{table:yse_stats} and sample light curves are shown in Figure~\ref{fig:real_lcs}.  Magnitude and redshift histograms for YSE transients to date are shown in Figure~\ref{fig:inithists}.

To date, YSE observations have a median PSF full width at half maximum (FWHM) of 1.3 arcsec and 5-$\sigma$ detection limits of $gri \approx 21.5, 21.7, 21.4$~mag during dark time and $riz \approx 20.9, 20.9, 20.5$~mag during bright time (Table~\ref{table:yse_overview}).  After weather loss, we observe with a median cadence of 3.9~days.  The median phase of first ${\rm S/N} > 3$ observation for YSE transients (based on estimates of the time of maximum light) is $-6.4$~days\footnote{This phase is roughly equivalent to the discovery epoch, but our discovery reports are sometimes delayed as discussed in Section~\ref{sec:reporting}.}.

\begin{figure}
    \centering
    \includegraphics[width=3.4in]{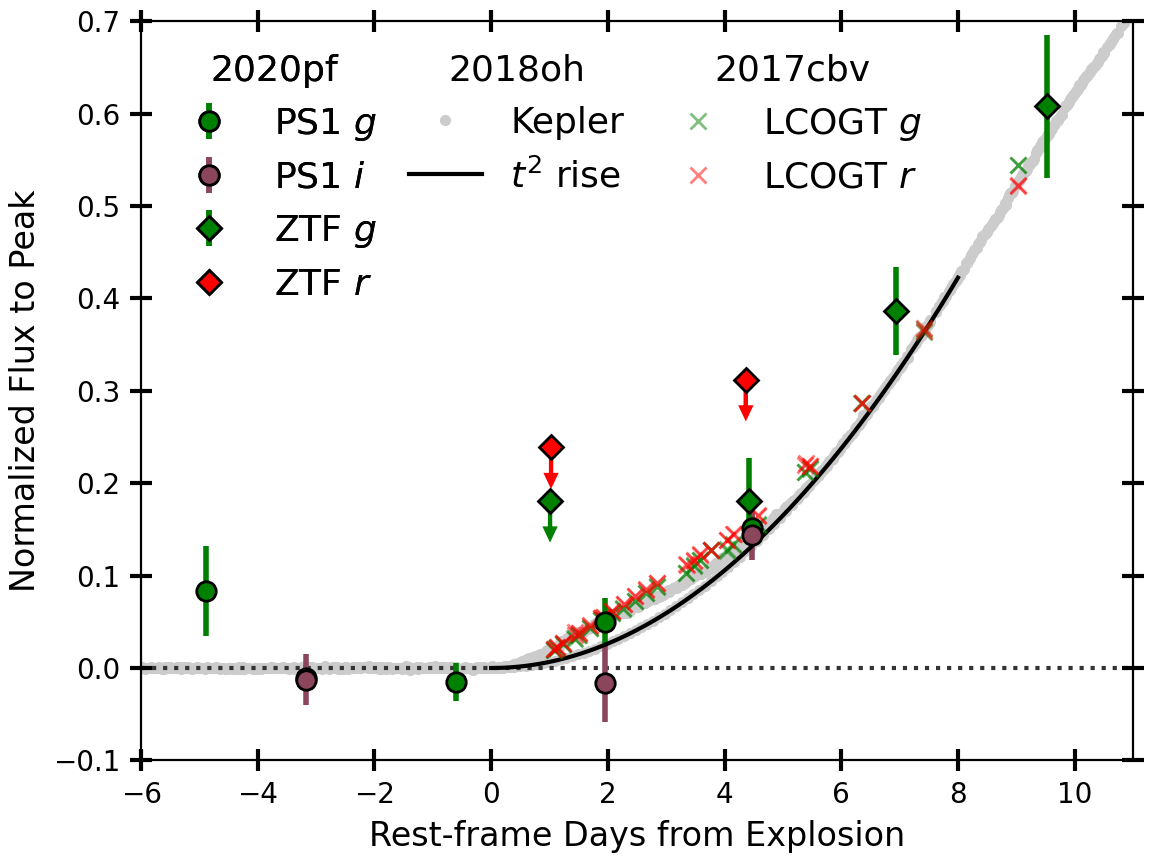}
    \caption{YSE (circles) and ZTF (diamonds) $gri$ light curve of the relatively distant ($z = 0.095$) SN~Ia~2020pf covering the time of explosion.  We detect SN~2020pf 2~days after explosion.  For comparison, we show the best-fit $t^{2}$ rising light curve as well as early flux-excess SNe~Ia~2017cbv and 2018oh \citep[crosses and grey points, respectively;][]{Hosseinzadeh17, Dimitriadis19}.}
    \label{fig:2020pf}
\end{figure}

YSE observations include \nspecclass\ spectroscopically classified transients (Figure~\ref{fig:class_pie}); highlights include two SNe~Iax (2020inp and 2020sck), one super-Chandrasekhar SN\,Ia (2020esm), four SNe~IIb (2020fqv, 2020ikq, 2020ivg, 2020tkc), one SN~Ic-BL (2020fhj), and three TDEs: 2020neh, 2020nov, and 2020opy, with AT~2020neh being particularly unusual in its rapid evolution\footnote{See Astronomer's Telegrams and AstroNotes from YSE team membors for 2020tlf and 2020tkc \citep{Dimitriadis20ATel}, 2020fhj \citep{Izzo20ATel}, 2020inp \citep{Dimitriadis20ATelb}, 2020neh \citep{Angus20ATel}.}.  We have observed several SNe~Ia within 2--3 days of explosion, including SNe~2020fci, 2020ioz, 2020juq, 2020nbo, and 2020pf, among other candidates.  We also observed flash ionization features in Keck spectra of CC~SNe~2020pni and 2020tlf and, for 2020pni, we observed a strong UV peak in {\it Swift} photometry at approximately 1.9~days after explosion thanks to ZTF detections within one day of explosion and deep YSE non-detections two days prior (Figure~\ref{fig:20pni}).  The early rise of SN~Ia~2020pf is shown in Figure~\ref{fig:2020pf}, with the first detection occurring at approximately 19~days before B-band maximum light as determined from a fit of the light curve using SALT2.

We have obtained targeted YSE survey observations of nearby stripped-envelope SNe~2019yvr (Ib), 2020oi (Ic), and 2020nxt (Ibn) as well as four very nearby SNe~Ia, SNe~2020jgl and 2020uxz and Virgo-adjacent SNe~2020ue and 2020nlb; analyses of these exciting objects are forthcoming. SNe~2020ees, 2020fqv, and 2020nxt have contemporaneous {\it HST} observations, while an additional ten YSE transients have {\it HST} pre-explosion imaging. We are working with ongoing programs such as ``Supernovae in the near-Infrared avec Hubble'' (SIRAH; {\it HST}-GO 15889, PI: Saurabh Jha) to share promising young supernovae for coordinated observations with {\it HST} and other facilities. We welcome external collaborators; our external scientist policy together with a guide to the application can be found at \url{https://yse.ucsc.edu/collaborate/}.

Given telescope down time, we have effectively observed $\sim$7~months to date, which equates to approximately 3.5~months of the survey simulations in Section~\ref{sec:sim_yields} as they assume double the area coverage.  By this estimation we are slightly under-performing our expected yields of $\sim$5000 SNe per year, but this is expected as early survey photometry has overestimated photometric uncertainties by factors of 1.7--1.9 (see Section~\ref{sec:sim_yields}).  Our current magnitude-limited sample of 125 SNe is much larger than expected from the 108 predicted SNe per year in our simulations, though a handful of targeted SNe and survey edge effects as we catch slow-declining SNe when we begin observations of new fields have likely boosted this number significantly.  Our volume-limited sample also exceeds that expected from simulations, though this number may also be impacted by photo-$z$ uncertainties.

Our team aims to obtain a spectrum of every unclassified transient with peak $r < 18.5$~mag, $D < 250$~Mpc, or a detection within two days of explosion for an estimated total of 217 per year (Figure~\ref{fig:surveysims_magvol}).  
Our sample will necessarily include classifications by the community and we are grateful to the teams responsible for helping to classify transients in the YSE magnitude- and volume-limited samples.  To date, there have been significant contributions to classifications in this sample from ZTF, ePESSTO$+$ \citep{Smartt15}, and the SIRAH team.  Our secondary priority will be to spectroscopically classify transients that appear unusual based on their photometric properties.  

Our current magnitude-limited sample is 91\% spectroscopically complete.  We examined the failures and the two most common reasons for a lack of a spectrum are lack of spectroscopic resources because of COVID-19 shutdowns and the transient being discovered shortly before it sets.  Therefore, we are optimistic that our completeness can increase as the survey continues.

\section{Conclusions}
\label{sec:conclusions}

YSE is a transient survey that will use 7\% of the time on both the Pan-STARRS1 and Pan-STARRS2 telescopes to survey 1512 deg$^{2}$ in $griz$ with a 3-day cadence and two filters per epoch.  YSE began on November 24th, 2019, and has been running for the past several months using only Pan-STARRS1 but will soon double its observing time by using Pan-STARRS2.  Detailed simulations indicate that at full capacity, YSE will discover $\sim$5000 SNe per year.  YSE images reach limiting magnitudes of $\sim$21.5 in $gri$ and $\sim$20.5 in $z$, with dark- and bright-time observations taken in combinations of $gri$ and $riz$, respectively.  The YSE survey strategy avoids the Galactic plane and prioritizes equatorial fields with a high volume of archival and, when possible, multi-wavelength data.  YSE is currently planning to carry out a three-year survey; survey updates and contact information for members of our team can be found at \url{https://yse.ucsc.edu}.  Given that we have not yet begun surveying the full YSE area, we are optimistic that YSE will continue through the anticipated start of operations at Rubin Observatory in 2023.

The primary science drivers of YSE include opening a new discovery space for faint and red transients, building a census of transients in the nearby universe, understanding black hole variability and tidal disruption events, and assembling a legacy high-cadence, low-$z$ SN~Ia cosmology sample to complement high-redshift samples from the Rubin Observatory, and the {\it Roman Space Telescope}.  YSE has discovered or observed \ntotalsne\ SNe to date and \percentdisc\% of the likely transients announced so far in 2020.  

When Pan-STARRS2 observations commence, we anticipate that the YSE survey volume will be approximately $\sim$20\% that of ZTF and the $i$-band volume will be $\sim$75-90\% that of ZTF (Table \ref{table:yse_comp}; estimates depend on the luminosity of the transient).  We will have YSE data for approximately 10\% of ZTF transients in the public survey, giving thousands of light curves with combined observations every 1--2 days.  We will also have a sample of thousands of cosmologically useful SNe\,Ia with $\sim$3~mmag photometric calibration and be able to detect young SNe $\sim$1 day before they are detectable by other all-sky time-domain surveys.  The over-sampled Pan-STARRS PSF may make it easier to discover transients at the centers of galaxies.

The Young Supernova Experiment's datasets are yielding new insights into transient physics and are complementary to other ongoing surveys. YSE's multi-wavelength light curves and our team's follow-up spectroscopy are forming a rich dataset that will improve our understanding of the time-domain universe and prepare the community for the forthcoming Rubin Observatory and its Legacy Survey of Space and Time.

\acknowledgements
We thank S.\ Jha, R.\ Kirshner, K.\ Maguire, A.\ Riess, B.\ Schmidt, and D.\ Scolnic for discussion about survey strategy.  We thank J.\ Nunez for help and useful discussions involving the YSE-PZ transient management system and E.\ Strasburger and J.\ Johnson for help with spectroscopic follow-up observations.  We also thank the anonymous referee for their helpful suggestions to improve the manuscript.

D.O.J.\ acknowledges support from a Gordon and Betty Moore Foundation postdoctoral fellowship at the University of California, Santa Cruz.  D.O.J, K.D.A and K.D.F. also acknowledge support provided by NASA Hubble Fellowship grants HST-HF2-51462.001, HST-HF2-51403.001, and HST-HF2-51391.001-A, respectively, which are awarded by the Space Telescope Science Institute, operated by the Association of Universities for Research in Astronomy, Inc., for NASA, under contract NAS5-26555.  The UCSC team is supported in part by NASA grants NNG17PX03C, 80NSSC19K1386, and 80NSSC20K0953; NSF grants AST-1518052, AST-1815935, and AST-1911206; the Gordon \& Betty Moore Foundation; the Heising-Simons Foundation; and by a fellowship from the David and Lucile Packard Foundation to R.J.F.  D.A.C.\ and M.R.S.\ are supported by the National Science Foundation Graduate Research Fellowship Program Under Grant Nos.\ DGE-1339067 and DGE-1842400, respectively. E.R.-R.\ is supported by the  Heising-Simons Foundation and NSF (AST-1852393 and AST-1911206).  E.R.-R.\ and H.P.\ are supported by the Danish National Research Foundation (DNRF132).  H.P.\ is also indebted to the Hong Kong government (GRF grant HKU27305119) for support.

W.J.-G.\ is supported by the National Science Foundation Graduate Research Fellowship Program under Grant No.~DGE-1842165 and the IDEAS Fellowship Program at Northwestern University. C.D.K. acknowledges support through NASA grants in support of {\it Hubble Space Telescope} program AR-16136.  The computations in this paper were aided by the University of Chicago Research Computing Center.   RM acknowledges partial support by the National Science Foundation under Award No. AST-1909796 and AST-1944985 and by the Heising-Simons Foundation under grant \#~2018-0911. RM is a CIFAR Azrieli Global Scholar in the Gravity \& the Extreme Universe Program, 2019 and a Alfred P. Sloan Fellow in Physics, 2019.

This work was supported by a VILLUM FONDEN Investigator grant to J.H. (project number 16599) and two Villum Young Investor Grants to C.G. and C.G. (10123 and 25501).  

G.N. and K.D.F are grateful for support from the University of Illinois at Urbana-Champaign. D.C. is supported by the Center for Astrophysical Surveys at NCSA through the Illinois Survey Science Postdoctoral Fellowship. A.G. is supported by the National Science Foundation Graduate Research Fellowship Program under Grant No. DGE – 1746047. A.G. and P.D.A. are supported by the Center for Astrophysical Surveys as Illinois Survey Science Graduate Fellows. A.G.\ additionally acknowledges support from the Illinois Distinguished Fellowship. A.E.'s work was supported by a 2019 Physics \& Astronomy Student Travel Grant from the Council for Undergraduate Research. The computations in this paper made extensive use of the Illinois Campus Cluster, and facilities at the National Center for Supercomputing Applications at UIUC. 

Parts of this research were supported by the Australian Research Council Centre of Excellence for All Sky Astrophysics in 3 Dimensions (ASTRO 3D), through project number CE170100013.

S.J.S.\ and K.W.S.\ are supported by STFC grants ST/P000312/1, ST/S006109/1 and ST/T000198/1.

A.H.\ is supported by Future Investigators in NASA Earth and Space Science and Technology (FINESST) award No.\,80NSSC19K1422.

M.R.D. acknowledges support from the NSERC through grant RGPIN-2019-06186, the Canada Research Chairs Program, the Canadian Institute for Advanced Research (CIFAR), and the Dunlap Institute at the University of Toronto.

This project has received funding from the European Union’s Horizon 2020 research and innovation programme under the Marie Sklodowska-Curie grant agreement No 891744 to S.I.R.  Additionally S.I.R.\ gratefully acknowledges support from the Independent Research Fund Denmark via grant numbers DFF 4002-00275 and 8021-00130.

The Pan-STARRS1 Surveys (PS1) and the PS1 public science archive have been made possible through contributions by the Institute for Astronomy, the University of Hawaii, the Pan-STARRS Project Office, the Max-Planck Society and its participating institutes, the Max Planck Institute for Astronomy, Heidelberg and the Max Planck Institute for Extraterrestrial Physics, Garching, The Johns Hopkins University, Durham University, the University of Edinburgh, the Queen's University Belfast, the Harvard-Smithsonian Center for Astrophysics, the Las Cumbres Observatory Global Telescope Network Incorporated, the National Central University of Taiwan, the Space Telescope Science Institute, the National Aeronautics and Space Administration under Grant No. NNX08AR22G issued through the Planetary Science Division of the NASA Science Mission Directorate, the National Science Foundation Grant No.\ AST-1238877, the University of Maryland, Eotvos Lorand University (ELTE), the Los Alamos National Laboratory, and the Gordon and Betty Moore Foundation.

Some of the data presented herein were obtained at the W.\ M.\ Keck Observatory, which is operated as a scientific partnership among the California Institute of Technology, the University of California and the National Aeronautics and Space Administration. The Observatory was made possible by the generous financial support of the W.\ M.\ Keck Foundation.  The authors wish to recognize and acknowledge the very significant cultural role and reverence that the summit of Maunakea has always had within the indigenous Hawaiian community.  We are most fortunate to have the opportunity to conduct observations from this mountain.

Based on observations obtained with the Samuel Oschin 48-inch Telescope at the Palomar Observatory as part of the Zwicky Transient Facility project. ZTF is supported by the National Science Foundation under Grant No. AST-1440341 and a collaboration including Caltech, IPAC, the Weizmann Institute for Science, the Oskar Klein Center at Stockholm University, the University of Maryland, the University of Washington, Deutsches Elektronen-Synchrotron and Humboldt University, Los Alamos National Laboratories, the TANGO Consortium of Taiwan, the University of Wisconsin at Milwaukee, and Lawrence Berkeley National Laboratories. Operations are conducted by COO, IPAC, and UW.

We acknowledge the use of public data from the {\it Neil Gehrels Swift Observatory} data archive.

This research has made use of the NASA/IPAC Extragalactic Database (NED), which is operated by the Jet Propulsion Laboratory, California Institute of Technology, under contract with the National Aeronautics and Space Administration.

\appendix
\label{sec:appendix}

\section{YSE Survey Simulation Variants and Methodology}
\label{sec:sim_appendix}

To estimate SN discovery statistics from YSE, we used the SNANA simulation software \citep{Kessler10}.  SNANA generates catalog-based simulations that use real survey noise properties, detection efficiencies, SN rates, and luminosity functions, as well as the transient SED models used for the PLAsTiCC SN classification challenge \citep{Kessler19,Malz19}.  We base the YSE simulations on simulations that were originally generated for the Foundation Supernova Survey and used for the cosmological parameter measurements in \citet{Jones19}.  Because Foundation took predominantly 15-second exposures, we were able to generate an accurate YSE survey realization simply by scaling the exposure times to match the 5-$\sigma$ detection limits shown in Figure~\ref{fig:depth}.  This may introduce small errors in the relation between S/N and brightness, but these should be negligible as our analysis focuses on the number of detected SNe in the simulations (with the exception of cosmologically useful SNe~Ia in Table~\ref{table:yse_strategy}, which may scale up or down slightly depending on the true survey noise properties).

The PLAsTiCC transient SED models are described in detail in \citet{Kessler19}.  We do not simulate any variable sources (e.g., stars and AGN) or purely theoretical models (e.g., pair-instability SNe or cosmic strings).  However, we simulate the full range of available SN~Ia and CC~SN subtypes, as well as SNe~Iax, Ca-rich transients, and intermediate luminosity optical transients (ILOTs).  The PLAsTiCC models come from a wide variety of sources; we note that at early times in particular, they are based on scant observational data.

For the simulated SN rates, we again follow PLAsTiCC, who use SN~Ia rates from \citet{Dilday08}:
\begin{equation}
    R_{\rm Ia} (z) = 2.5 \times 10^{-5} (1+z)^{1.5}~{\RateUnit} ~~(z<1),
\end{equation}
while CC~SN rates are from \citet{Strolger15}:
\begin{equation}
    R_{\rm CC} (z) = 5.0 \times 10^{-5} (1+z)^{4.5}~{\RateUnit} ~~(z<1).
\end{equation}
In most cases, normalization of the rates for different CC~SN and SN~Ia subtypes are given by the volume-limited measurements of \citet{Li11}, but \citet{Kessler19} lists a few exceptions for peculiar events.  \citet{Kessler19} also describes the adopted redshift-dependent rates of peculiar classes of events.  While the PLAsTiCC models and SNANA simulations suffer from a number of uncertainties in both the underlying models and in the rates and luminosity functions of both rare and relatively common SN subtypes, PLAsTiCC is the most reliable currently available compilation combining rates, luminosity functions, and spectrophotometric transient models (however, see \citealp{Vincenzi19} for a recent update to a number of CC~SN templates).

As a consistency check, we integrate these rates over our expected 1512~deg$^{2}$ survey area and within volumes where we would expect to discover nearly every SN given the survey depth.  Assuming 24\% loss for detector masking and an additional 30\% for weather, we predict a discovery rate of $\sim$2300 SNe~Ia per year to a distance of 1~Gpc and 76 CC~SNe per year to a distance of 250~Mpc, distances corresponding to where a 27-second PS1 exposure should be sufficiently deep to detect nearly all SNe~Ia and CC~SNe.  Our baseline simulation predicts that we should detect 1840 SNe~Ia per year within 1~Gpc and 116 CC~SN per year within 250~Mpc, consistent with these estimated numbers, with the increase in CC~SNe likely due to the longer duration of many of these events.

We conservatively assume 24\% masking of the GPC1 and GPC2 cameras, which will likely give a slight underestimate of the true SN yields and the area shown in Table \ref{table:yse_strategy} includes these losses.  Per-month estimates of weather loss are taken from PS1 data from the summit of Haleakala between 2010 and 2014 and are directly included in the simulation to account for correlations in downtime.  We also varied the simulated survey depth as a function of lunar phase to match what we observe from Foundation data and the 3$\pi$ survey; $gri$ depths are estimated to vary by approximately 1.0, 0.7, and 0.2~mag between dark time and bright, while $z$ depths are assumed to remain constant.  Typical night-to-night variation in depth is included.  We also simulated an approximate ZTF survey by assuming an average depth of 21.1~mag in dark time that again varies to the same degree with moon phase as PS1 and we assumed that ZTF observations (when simulated Palomar and Haleakala weather allows) will always take place 21 hours after YSE observations.  ZTF weather is estimated from monthly averages of weather loss at Palomar from the first year of ZTF observations.

A small number of second-order effects are not included in these simulations.  First, we do not include loss due to the Pan-STARRS moon avoidance angle (30-35$^{\circ}$) in the simulations.  Because we prioritize fields far from the ecliptic, this loss can usually be limited to approximately five days per month depending on the field.  We also do not simulate the changing of the telescope position angle, which changes the location of masked pixels on the sky and is effectively a 20\% weather-like reduction in the sampling of light curves (but {\it not} the area on sky).  Finally, though fields will typically be observed for approximately 4--6 months per year, we do not include edge effects from rising or setting fields.  We also do not simulate the Virgo one-day survey.  However, none of these effects should cause substantial reductions in the numbers or demographics of simulated SNe, which are our primary goals in this section.

The nominal survey strategy along with several alternate strategies considered by our team are shown in Table~\ref{table:yse_strategy}.  We caution that some of the numbers, particularly discoveries of young SNe, are affected by small-number statistics.  Due to the limited filter choices during bright time to avoid the drastic reduction in $g$-band depth, we simulated a $ri+rz$ filter sequence during bright time for all two-filter survey designs.  We found that as exposure time increases, the predicted number of SN discoveries also tends to increase but begins to flatten after 25s, which is unsurprising given the volume as a function of exposure time shown in Figure \ref{fig:depth}.  Choosing lower exposure times increases the number of transient discoveries for which we can obtain spectroscopic follow-up observations, while higher exposure time increases the total number of transients.  We decided on a survey with two filters per epoch because the survey designs featuring three and four filters per epoch would be very valuable for some science cases but would substantially reduce the transient discovery volume.  Finally, a blue-focused survey would probe a more similar discovery space to other ongoing surveys such as ZTF, while a red-focused survey would be expected to reduce the number of young SNe discovered as these are expected to be blue.  For these reasons, we chose a two-filter survey with a $gr+gi$ filter sequence during bright time with exposure times in each band of 27~seconds.

\begin{table*}
\caption{\fontsize{9}{11}\selectfont Survey Strategy Comparisons for YSE}
  \centering
\begin{tabular}{lrrrrrrrrrrrrr}
  \hline \hline\\[-1.5ex]
&\multicolumn{13}{c}{Nominal Strategy}\\*[2 pt]
&\multicolumn{3}{c}{Survey Parameters}&&\multicolumn{3}{c}{SN Discoveries Per Year}&&\multicolumn{2}{c}{Young SNe Per Year}&&\multicolumn{2}{c}{Ia Cosmology$^{\mathrm{a}}$}\\
Filters$^{\mathrm{b}}$&Exp. (s)&N$_{\mathrm{filt}}$/day&Survey Area$^{\rm c}$&&$z_{\mathrm{med}}$&N$_{\mathrm{Ia}}$&N$_{\mathrm{CC}}$&&N($\le$2 d)&N($\le$3 d)&&$z_{\mathrm{med}}$&N$_{\mathrm{Ia}}$/yr\\
  \hline\\[-1.5ex]

$gr+gi$&27&2&1149 deg$^2$&&0.19&4088&1063&&4&28&&0.12&812\\
\\[-1ex]
\tableline\\*[2 pt]
&\multicolumn{13}{c}{Varying Exposure Times}\\*[2 pt]
&\multicolumn{3}{c}{Survey Parameters}&&\multicolumn{3}{c}{SN Discoveries Per Year}&&\multicolumn{2}{c}{Young SNe Per Year}&&\multicolumn{2}{c}{Ia Cosmology}\\
Filters&Exp. (s)&N$_{\mathrm{filt}}$/day&Survey Area&&$z_{\mathrm{med}}$&N$_{\mathrm{Ia}}$&N$_{\mathrm{CC}}$&&N($\le$2 d)&N($\le$3 d)&&$z_{\mathrm{med}}$&N$_{\mathrm{Ia}}$/yr\\
\hline \\*[-1.5ex]
$gr+gi$&15&2&1679 deg$^2$&&0.16&3463&817&&2&25&&0.11&759\\
$gr+gi$&20&2&1409 deg$^2$&&0.17&3700&910&&0&24&&0.11&760\\
$gr+gi$&25&2&1213 deg$^2$&&0.19&4119&1050&&3&34&&0.12&841\\
$gr+gi$&30&2&1065 deg$^2$&&0.20&4151&1061&&8&35&&0.13&843\\
$gr+gi$&35&2&949 deg$^2$&&0.21&4057&1092&&3&25&&0.14&770\\
\\[-1ex]
\tableline\\*[2 pt]
&\multicolumn{13}{c}{Varying Number of Filters}\\*[2 pt]
&\multicolumn{3}{c}{Survey Parameters}&&\multicolumn{3}{c}{SN Discoveries Per Year}&&\multicolumn{2}{c}{Young SNe Per Year}&&\multicolumn{2}{c}{Ia Cosmology}\\
Filters&Exp. (s)&N$_{\mathrm{filt}}$/day&Survey Area&&$z_{\mathrm{med}}$&N$_{\mathrm{Ia}}$&N$_{\mathrm{CC}}$&&N($\le$2 d)&N($\le$3 d)&&$z_{\mathrm{med}}$&N$_{\mathrm{Ia}}$/yr\\
\hline \\*[-1.5ex]
$gri+riz$&27&3&766 deg$^2$&&0.20&3059&778&&3&23&&0.14&783\\
$griz$&27&4&575 deg$^2$&&0.20&2367&643&&1&13&&0.15&649\\
\\[-1ex]
\tableline\\*[2 pt]
&\multicolumn{13}{c}{Blue/Red Strategies}\\*[2 pt]
&\multicolumn{3}{c}{Survey Parameters}&&\multicolumn{3}{c}{SN Discoveries Per Year}&&\multicolumn{2}{c}{Young SNe Per Year}&&\multicolumn{2}{c}{Ia Cosmology$^{\mathrm{a}}$}\\
Filters&Exp. (s)&N$_{\mathrm{filt}}$/day&Survey Area&&$z_{\mathrm{med}}$&N$_{\mathrm{Ia}}$&N$_{\mathrm{CC}}$&&N($\le$2 d)&N($\le$3 d)&&$z_{\mathrm{med}}$&N$_{\mathrm{Ia}}$/yr\\
\hline \\*[-1.5ex]
$gr$&27&2&1149 deg$^2$&&0.19&3960&1042&&2&28&&0.11&534\\
$ri+rz$&27&2&1149 deg$^2$&&0.20&4287&1128&&6&20&&0.12&20$^{\rm d}$\\

\hline \\[-1.5ex]

\multicolumn{14}{l}{
\begin{minipage}{17cm}
\begin{itemize}[wide,labelindent=0pt]
    \item[$^{\mathrm{a}}$] Number of predicted cosmologically useful SNe~Ia after applying standard selection criteria \citep{Scolnic18} and requiring total distance modulus uncertainty (including intrinsic dispersion) $<0.15$~mag.
    \item[$^{\mathrm{b}}$] These filter combinations are for dark time; in bright time, we simulate combinations of $ri$ and $rz$ for all two-filter survey strategies due to limited filter options that would achieve reasonable depths.  In the case of three and four filters per day, the same strategies are used for both bright and dark time observations.  The ``$gr$+$gi$'' sequence indicates that our survey design alternates these two filter combinations with the full sequence repeating every six days.
    \item[$^{\rm c}$] Area {\it after} assuming 24\% masking of the GPC detectors per epoch.
    \item[$^{\rm d}$] The small number is largely because SALT2 does not include the rest-frame $iz$ bands in light-curve fitting.
\end{itemize}
\end{minipage}
}
\end{tabular}
\label{table:yse_strategy}
\end{table*}

\section{Components of the Field Selection Metric}
\label{sec:fields_appendix}

\begin{figure*}
    \centering
    \includegraphics[width=7in]{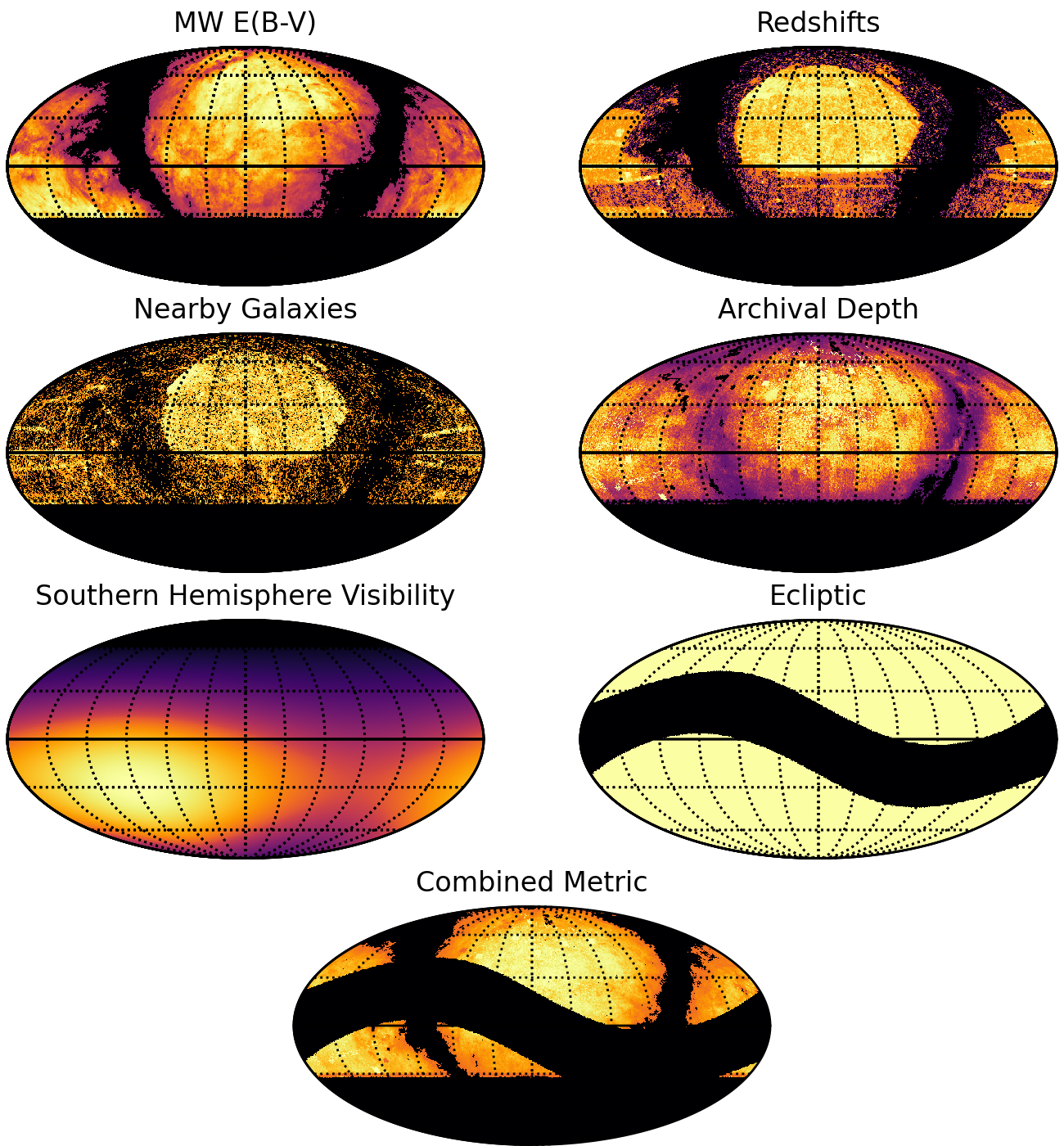}
    \caption{Components of the YSE field selection metric (top six panels) and combined metric map (bottom panel) with lighter colors corresponding to higher values of the metric.  We note that a number of the input maps shown here are restricted to the 3$\pi$ coverage area ($\delta > -30^{\circ}$).  We also prioritize the amount of {\it HST} exposure time, particularly interesting galaxies/clusters, and fields with newly discovered SNe from other surveys, but do not formally include these in the metric.  High-priority fields that do not pass our cut on proximity to the ecliptic are sometimes included in YSE.}
    \label{fig:metric_appendix}
\end{figure*}

The total YSE field selection metric $m$ is computed from the following equation:

\begin{equation}
    m = m_{\rm MW} \times m_{d} + m_{z} + m_{\rm DECaLS} + m_{ng} + m_f + m_e.
\end{equation}

\noindent The metric is computed from individual metrics. including the Milky Way $E(B-V)$ from \citet{Schlafly11} ($m_{\rm MW}$), the combined depths from SDSS, Pan-STARRS 3$\pi$ and MDS ($m_{d}$), the number of galaxies with measured redshifts  ($m_z$), coverage from DECaLS ($m_{\rm DECaLS}$; \citealp{Dey19}), the number of nearby galaxies $m_{ng}$, follow-up capability from Chile (m$_f$), and proximity to the ecliptic plane (m$_e$).

Each of these individual metrics is parameterized and scaled to ensure reasonable weights and a final range from 0 to 1.  $m_{\rm MW}$ is set to zero for MW $E(B-V)$ greater than 0.2 and is proportional to the inverse extinction $[1+E(B-V)]^{-1}$.  The depth metric $m_{d}$ uses $r$-band depths from 20.5 ($m_d = 0$) to 22.5~mag ($m_d = 1$) scaled linearly; fields with PS1 MDS coverage are given a metric of 1, and the depth of either 3$\pi$ or SDSS for non-MDS regions of sky is used depending on which is deeper (typically PS1, except for Stripe 82).  For the nearby galaxies metric $m_{ng}$, we weight by the number of galaxies at $<150$, $<40$, $<20$, and $<10$~Mpc from the NASA/IPAC Extragalactic Database (NED) and combine each of these sets; e.g., galaxies at $<10$~Mpc contribute to the number of galaxies in all four maps and therefore have four times the effective weight of 150 Mpc galaxies. 
The nearby galaxies metric is added to the redshifts metric $m_z$, which is the number of redshifts from the combination of SDSS and NED.  The resulting map is clipped at the 98th percentile to reduce the influence of large outliers.
The DECaLS depth metric $m_{\rm DECaLS}$ is computed in the same manner as $m_{d}$ but given half the weight as the other surveys, and not multiplied by the MW E(B-V), given its predominantly $z$-band coverage.  The follow-up observation metric $m_{f}$ is arbitrarily determined as the product of the airmass and hours of observation per night from La Silla observatory in Chile averaged over a year.  Most of the high-metric fields are in the Northern hemisphere and this choice weights YSE survey fields for which spectroscopic follow-up observations from Chile are possible.  Finally, unless a field is of particular scientific interest we exclude those fields within 20~deg of the ecliptic plane; due to the Pan-STARRS moon avoidance angle of 30~deg, fields near the ecliptic may have 8-10~day gaps in their light curves.  The metric map is then normalized between 0 and 1.

The combination of individual metric quantities along with the final map are shown in Figure~\ref{fig:metric_appendix}.

\bibliographystyle{apj}
\bibliography{main}

\allauthors
\end{document}